\documentclass[a4paper]{article}
\usepackage{bm,amsmath,amsfonts}
\usepackage{natbib}

\newcommand{\ket}[2]{|#1\rangle _{#2}}
\newcommand{\bra}[2]{\langle _{#1}#2|}
\newcommand{\up}{\!\uparrow}

\title{Quantum Bayesianism: A Study}
\author{Christopher G. Timpson\thanks{\texttt{christopher.timpson@bnc.ox.ac.uk}} \\ \textit{\small Brasenose College,}\\ \textit{\small University of Oxford, OX1 4AJ, UK}} 

\date{27 September 2007}

\begin{document}
\maketitle

\begin{abstract}
\noindent The Bayesian approach to quantum mechanics of Caves, Fuchs and Schack is presented. Its conjunction of realism about physics along with anti-realism about much of the structure of quantum theory is elaborated; and the position defended from common objections: that it is solipsist; that it is too instrumentalist; that it cannot deal with Wigner's friend scenarios. Three more substantive problems are raised: Can a reasonable ontology be found for the approach? Can it account for explanation in quantum theory? Are subjective probabilities on their own adequate in the quantum domain? The first question is answered in the affirmative, drawing on elements from Nancy Cartwright's philosophy of science. The second two are not: it is argued that these present outstanding difficulties for the project. A quantum Bayesian version of Moore's paradox is developed to illustrate difficulties with the subjectivist account of pure state assignments.
\end{abstract} 

\newpage

\tableofcontents

\newpage

\begin{quote}\small ``Covering-law theorists tend to think that nature is well-regulated; in the extreme that there is a law to cover every case. I do not. I imagine that natural objects are much like people in societies. Their behaviour is constrained by some specific laws and by a handful of general principles, but it is not determined in detail, even statistically. \textit{What happens on most occasions is dictated by no law at all.}" \citep[p.49]{cartwright:hlpl} 
\end{quote}

\section{Introduction}

The development of the field of quantum information theory has brought along with it a welcome boost to work in the foundations of quantum theory. Old questions have been revisited (e.g., `What is the nature of entanglement?', `What is the nature of non-locality?', `How are measurement and disturbance \textit{really} related?') and new mathematical tools and new concepts have been brought to bear; while many entirely new questions have been posed. Foremost amongst these, perhaps, are the questions of whether quantum information theory might finally provide the resources to resolve our longstanding conceptual difficulties with quantum mechanics; and whether it might be possible to \textit{derive} quantum theory from information-theoretic principles. Very different possibilities for information processing are available according to classical and to quantum theories. Perhaps understanding these differences---and further differences from wider classes of theories in which restrictions on information processing differ yet again (cf. \citet{barrett:generalized,bblw:broadcasting})---will help us to understand \textit{why} we have a theory structured as quantum theory is.\footnote{\citet{timpson:paqit} may be consulted for an introduction to the ideas of quantum information theory, along with further references.}

Within this broad programme of seeking to understand quantum mechanics with the aid of resources from quantum information theory, one of the most interesting---and radical---proposals to date is what can be called the \textit{quantum Bayesianism} of Caves, Fuchs and Schack \citep{fuchs:only,cfs:deF,fuchs:what,cfs:compatibility,cfs:certainty}. Fuchs has been at the forefront of the call for information-theoretic axiomatisation of quantum theory \citep{fuchs:paulian}. He has urged that for each of the familiar components of the abstract mathematical characterisation of quantum mechanics (states are represented by density operators, they live on a complex Hilbert space, etc.) we should seek to provide an information-theoretic reason, if possible. The thought is that once a \textit{transparent} axiomatisation of quantum mechanics has been achieved, we will have available a conceptual framework for thinking about the theory which will render it unmysterious, much as the meaning of the Lorentz transformations is rendered unmysterious by Einstein's Light Postulate and Principle of Relativity. Moreover, Fuchs urges that an interpretation of quantum mechanics should be judged on the extent to which it aids us in the completion of this project; the extent to which it helps us understand why we have quantum mechanics: why the theory is just as it is. It is the absence of any such account which he deems responsible for the (apparent) interminability of the arguments between proponents of the various different interpretations of quantum mechanics.

What is radical about the approach adopted by Caves, Fuchs and Schack is the starting point: it is to maintain that probabilities in quantum mechanics are subjective, across the board. This means that \textit{quantum states} will be subjective too: a matter of what degrees of belief one has about what the outcomes of measurement will be. States will not, then, be objective features of the world. Subjectivism, it is argued, is the natural way to understand probability; and it is the natural way, therefore, to understand quantum states. It allows, for example, a clean dissolution of traditional worries about what happens on measurement; and it dissolves worries about non-local action in EPR scenarios. Furthermore, it provides a novel way of approaching the business of axiomatisation. For if quantum states are subjective, features of agents' beliefs rather than features of the world, we may ask: what other components of the quantum formalism might be subjective too? If we can identify these and then whittle them away, we will be left with a better grasp of what the \textit{objective core} of the theory is.

Now, as a formal proposal, quantum Bayesianism is relatively clear and well developed. But it is rather less transparent philosophically. What exactly is at stake when one adopts this line? Is such an apparently radical approach sustainable? What would we have to be saying the world is like if quantum Bayesianism were the right way to understand it? It is the philosophical underpinnings of the approach which are the subject of this study. 

We will begin by setting the scene for the quantum Bayesian programme by outlining the interesting conjunction of realism and anti-realism that provides the distinctive rationale for the approach: a general realism about physics, combined with anti-realism about much of the structure of quantum mechanics (Section~\ref{scene}). Within such a setting, the quantum Bayesian's search for axioms for quantum theory takes on a special character: if it is supposed that no directly descriptive theory of the fundamental physical level is possible, then our attempts to grasp what there is at the fundamental level, and to understand how it behaves, will need to proceed indirectly. Next, more detail of the quantum Bayesian proposal is presented (Sections~\ref{outline},~\ref{detail}) and the important question broached of how, if quantum states are supposed to be subjective, scientists nevertheless tend to end up agreeing on what the right states for particular systems are. Further, the crucial issue of how the notoriously shifty divide between system and apparatus is to be managed in this approach is discussed (Section~\ref{apparatus}). It is suggested that the quantum Bayesian has rather less trouble here than one might suppose.

In Section~\ref{not solipsism}, three common objections to the proposal are aired; and then rebutted. The objections are these: that the approach is tacitly committed to solipsism; that it is overly instrumentalist; and that it cannot adequately deal with the data that would be empirically available in a Wigner's friend scenario. With these objections disposed of, the virtues of the quantum Bayesian programme are re-stated (Section~\ref{virtues}).

The ground is now clear to consider some more substantive challenges (Section~\ref{challenges}). Three in particular suggest themselves: 1) Can one find some kind of sensible ontology to underly the quantum Bayesian's approach? 2) Can one make requisite sense of explanations which involve quantum theory if one takes the Bayesian line? and 3) Are subjective probabilities in quantum theory really adequate? The first question is answered in the affirmative: it seems one can make out a suitable ontological picture, particularly if one makes use of components of Nancy Cartwright's philosophy of science, specifically, her emphasis on the causal powers or dispositions of objects as prior to any lawlike generalisations which might be true of them (cf. \citet{cartwright:dappled}).

The second two questions prove more problematic, however. By the quantum Bayesian's lights, much talk involving quantum mechanics will be non-descriptive: assignment of a state to a system will not involve characterising the properties of that system, for example. But then it becomes unclear how a large class of explanations which make use of quantum mechanics---those which proceed by explaining the properties of lager systems in terms of the properties possessed by their constituents and the laws governing them---could be supposed to function. \textit{Prima facie}, the quantum Bayesian approach would rob quantum theory of explanatory power which it nonetheless seems to possess (Section~\ref{explanation}). Regarding subjective probabilities (Section~\ref{subprobs}) it is argued that the quantum Bayesian is committed to a certain objectionable class of statements: what can be thought of as quantum versions of Moore's paradoxes (\citet[II.x]{wittgenstein:investigations}). These seem to betray something wrong in the links that the quantum Bayesian can allow between beliefs about outcomes which are certain to occur and the reasons which one could have for believing them to be so. Finally, the worry is expressed that the quantum Bayesian's subjectivism about probability renders mysterious how the means of enquiry about the world (gathering data by experiment) could deliver the intended ends: coping better with what the world has to throw at us. It may be that these concerns arising from questions (2) and (3) do not constitute insurmountable objections to the quantum Bayesian's position, but equally, it seems that they are concerns which cannot lightly be dismissed.

A final note before proceeding. The quantum Bayesian position to be discussed here is attributed jointly to Caves, Fuchs and Schack. It is clear that each is fully committed to the subjective Bayesian conception of probability and correlatively, of quantum states; the proposal has been developed collaboratively between them. But this leaves room for the possibility of some disagreement between them in matters of detail; and particularly when it comes to filling in the details of the philosophical package which might be developed to underpin the position. Here the primary textual sources are Fuchs' writings, important amongst which are his collections of email correspondence sent to colleagues and co-workers \citep{fuchs:paulian,fuchs:what,fuchs:cg}. The exposition that follows should therefore be understood as most closely informed by consideration of Fuchs' view, while still aiming to represent the views of Caves and Schack, but perhaps to a lesser extent. There may well be points---perhaps important ones---on which they would wish to demur, or to remain uncommitted. With this caveat in place, we may turn to setting the scene.

\section{Setting the Scene}\label{scene}

Let us begin with the straightforward thought that science in general---and physics in particular---is engaged in the business of finding out how things are and how things work. That the world we investigate exists and has it various physical characteristics independently of any of our thoughts and feelings about it; that finding out in science is a (joyfully rich!) development and extension of our common-sense means of finding things out in everyday life. 

Let us then grant that as physics progresses, we wish to develop theories that are not only increasingingly accurate in their predictions, but are also, at the same time, \textit{explanatory}. Theories that are (in some sense) at least approximately true; theories that describe (and, perhaps, if appropriate, unify) the underlying objects and processes giving rise to the phenomena of our interest. And let us grant that, historically, physics has, by and large, been reasonably successful in achieving these aims (the common epistemological objections notwithstanding).

But now suppose that this realist progression of explanatory, descriptive, theory construction eventually runs into difficulties. Suppose that, although applying just the same kinds of exploratory techniques, the same kinds of reasoning and the same kinds of approach to theory construction that have served so well in the past, one nonetheless ends up with a fundamental theory which is not descriptive after all; a theory which, one slowly comes to realise, has no direct realist interpretation; a theory whose statements are not apt to describe how things are. And let us suppose that this eventuality does not arise through any lack of effort or failure of imagination in theory construction; nor through want of computational ability; nor through any mere psychological or sociological inhibition. Perhaps it is just the case  that once one seeks to go beyond a certain level of detail, the world simply does not admit of any straightforward description or capturing by theory, and so our best attempts at providing such a theory do not deliver us with what we had anticipated, or with what we had wanted. 
A descriptive theory in any familiar sense is not to be had, perhaps, not even for creatures with greater cognitive powers and finer experimental ability than our own, for the world precludes it. The world, perhaps, to borrow Bell's felicitious phrase \citep{bell:speakable}, is \textit{unspeakable} below a certain level. What then for the realist? 

Just this provides the starting point for the quantum Bayesian approach to understanding quantum mechanics.

In the quantum Bayesian picture, quantum mechanics is the theory which, it is urged, should not be thought of in standard realist terms; either, for example, by being a realist about the quantum state (e.g., Everett, GRW) or by seeking to add further realist components to the formalism (e.g., hidden variable theories, modal interpretations, consistent histories). Rather, we are invited to recognise quantum mechanics as being the best we can do, \textit{given that the world will not admit of a straightforward realist description}. That best, it is suggested, is not a theory whose central theoretical elements---quantum states, measurements and general time evolutions---are supposed to correspond to properties or features of things and processes in the world; rather, it is a structure which is to be understood in broadly pragmatic terms: it represents our best means for \textit{dealing with} (that is, for forming our expectations and making predictions regarding) a world which turns out to be recalcitrant at a fundamental level; resistant to our traditional---and natural---descriptive desires.

But we need not give up our realism on account of this! Rather, the project must be transformed. Granted, our traditional realist descriptive project has been stymied: we are to take seriously the suggestion that quantum mechanics (with any of the paraphenalia of familiar realist interpretations of the formalism eschewed) is the best theory that one can arrive at. Better---and closer to the descriptive ideal---cannot be achieved, runs the thought. But if this is so then our realist desires must be served indirectly. One need not give up on the task of getting a handle on how the world is at the fundamental level just because no direct description is possible; one can seek an indirect route in instead. Quantum mechanics may not be a descriptive theory, we may grant, but it is a significant feature that we have been driven to a theory with just this characteristic (and unusual) form in our attempts to deal with and systematize the world. The structure of that theory is not arbitrary: it has been forced on us. Thus by studying in close detail the structure and internal functioning of this (largely) non-descriptive theory we have been driven to, and by comparing and contrasting with other theoretical structures, we may ultimately be able to gain \textit{indirect} insight into the fundamental features of the world that were eluding us on any direct approach; learn what the physical features \textit{are} that are responsible for us requiring a theory of just \textit{this} form, rather than any other. And what more could be available if the hypothesis that the world precludes direct description at the fundamental level is true? All we can do is essay this ingenious indirect approach.

Thus the essence of the quantum Bayesian position is to retain a realist view of physics and of the world whilst maintaining that our fundamental theory---quantum mechanics---should not itself receive a surface realist reading; while no simple-minded realist alternative is to be had either. The story that quantum mechanics has to tell about the world needs must, on this conception, be an indirect one. It is the bold positive part of the research programme to try to tell this story.

\subsection{An outline of the position}\label{outline}

With the general setting of the approach thus sketched, how does the quantum Bayesian position proceed in more detail? Considered as an interpretation of quantum mechanics, the characteristic feature of quantum Bayesianism is a point already mentioned above: its non-realist view of the quantum state. This takes a distinctive form: 
\begin{quote}
\textit{The quantum state ascribed to an individual system is understood to represent a compact summary of an agent's degrees of belief about what the results of measurement interventions on a system will be, \textbf{and nothing more}}.
\end{quote}
Two important points: unlike in many non-realist views of the state (e.g. \citet{ballentine,peres:book}) quantum states are assigned to individual systems, not just to ensembles; and most important of all, the probability ascriptions arising from a particular state assignment are understood in a purely subjective, Bayesian manner, in the mould of de Finetti \citep{deF:probabilisimo,deF:foresight} (see also \citet{ramsey,savage,jeffrey:real}). Then, just as with a subjective Bayesian view of probability, there is no right or wrong about what the probability of an event is, with the quantum Bayesian view of the state, there is no right or wrong about what the quantum state assigned to a system is.

This is a radical conception: what might motivate us to believe it? One set of reasons might derive from general convictions about the notion of probability. Making out a non-mysterious and non-vacuous notion of objective probability is notoriously controversial (the problems with frequency and propensity accounts are well known, for example); by adopting a subjectivist view of probability, where probabilities are analysed simply as agents' degrees of belief rather than objective quantities fixed by the world, one avoids these perplexities. It is widely held that in a physically deterministic world, probabilities would all have to be subjective anyway (at least single case ones), based on our ignorance of the relevant initial conditions, so it is often  supposed that, at best, only when fundamental theories posit stochastic properties at a fundamental level in their laws is there any scope for non-subjective probabilities. Quantum mechanics is typically seen as just such a theory providing objective probabilities (e.g. \citet{giere}). But the quantum Bayesian maintains that one can retain the clarity of subjectivist probabilities right across the board; and that this is conceptually advantageous. Their analysis results in showing that it is quite consistent even to take quantum probabilites to be subjective too.

A second set of reasons derives from the resolution one achieves of the standard perplexities of quantum mechanics; from the interpretive traction one gains on the standard problems of measurement and nonlocality by adopting this approach. These problems are not so much resolved as \textit{dissolved} in this setting; they don't arise in the first place. Thus the problem of measurement as it is standardly construed is the problem of explaining how measurement interactions end up presenting us with definite outcomes when the typical result of such an interaction is just that system and apparatus end up in an entangled state; or it is the problem of specifying exactly how, when and why \textit{collapse} (as opposed to unitary dynamics) occurs on measurement.

But in the quantum Bayesian picture these concerns are predicated on a false assumption: that the quantum states assigned to systems represent actual properties of those systems. They do not (it is maintained) instead representing beliefs about what the results of measurement interactions will be. The process of collapse doesn't imply a sudden change in the properties of a system; it does not correspond to a physical process at all, hence not to any \textit{mysterious} physical process in want of explanation. It is simply an updating of one's beliefs about what the results of future measurements on the system will be; an updating that occurs whenever one has data to update upon. Measuring apparatuses have definite properties at all times and, on interaction with a quantum system, will produce some particular outcome. On observation of that outcome, the experimentalist will revise the state that they assign to the initial system, using the standard rules. This process need be no more mysterious than the familiar one of conditionalising one's prior probability in some hypothesis on receipt of some data; replacing one's prior $p(h)$ with the posterior $p(h|d)$ when one observes that $d$ obtains\footnote{Fuchs expands on this comparison in detail in \citet[\S 6]{fuchs:only}. \citet{bub:qprobs} demurs, although not persuasively.}. Furthermore, since states only represent beliefs about what the results of future measurement outcomes will be, the fact that one might assign an entangled state to system and apparatus taken together does not mean that one must deny that the apparatus has definite properties, deny that it has its pointer pointing in some definite position following measurement; rather it just means that one's anticipations for the future involve certain non-factorisable probability distributions for joint measurements on system and apparatus taken together, which is quite another thing. (This may not be quite obvious: we shall return to this point in Section~\ref{apparatus}.)

A familiar way of making the measurement problem vivid is by appeal to a Wigner's friend scenario \citep{wigner:1961}. Here we imagine Wigner's experimentalist friend in a lab, about to perform, say, a Stern-Gerlach measurement in the $z$-direction on a spin-half system prepared in an eigenstate of spin in the $x$-direction. Wigner himself remains outside the lab. The experiment is run. What state should the friend assign to the system and apparatus? What state should Wigner assign? One normally argues: The friend, presumably, sees some definite outcome of the experiment so, we assume, assigns one of the pure product states corresponding to spin in some definite direction for the system, along with a definite pointer position recording that direction. Wigner, however, positioned outside the lab, unable to see any measurement result and considering the lab as a whole---friend included---to be a closed system (hence one subject to unitary Schr\"{o}dinger evolution) will ascribe an entangled state to the system and measuring apparatus; and perhaps even to the friend. But who is right?

The answer, of course, from the current perspective, is that neither is right and neither is wrong. There is no fact of the matter about what the states ascribed to things should be, these are wholly subjective judgements; thus it is quite consistent for agents in differing epistemic positions---one within the lab and one without, say---to assign different states to things. There is no tension between them. If Wigner strolls into the lab to see what the result is, \textit{then} he will update his beliefs and assign a product state; but there is no question of his friend hanging in limbo until Wigner does so. There is no relevant change in \textit{anything} physical when he does so; the only changes are internal to the agent ascribing the state. Given this lack of conflict between state assignments, no measurement problem arises. (There are some interesting further subtleties to the Wigner's friend scenario but we shall not have space to explore them here.)

The story about nonlocality runs much the same. Envisage an EPR-type scenario: two widely separated parties, call them now Alice and Bob, share an entangled state; for example, the singlet state of two spin-half systems. Alice performs a measurement in some direction on her system and gets an `up' outcome, say. What's happened to Bob's system? The usual story \citep{EPR} is that nonlocality is involved\footnote{At least if one is assigning states to individual systems, but is not eschewing collapse as Everett, for example, would.} : Alice's action has a nonlocal effect on Bob's system. Whereas before it had no definite spin property, being in a(n improper) maximally mixed state, following her measurement, it jumps into a definite spin state, now having spin down. But on the quantum Bayesian picture, there is of course no such result of collapse. The consequence of Alice's measurement is just that she updates her state assignment to the two systems, now assigning them an anti-correlated product state rather than the entangled singlet state. But that change doesn't correspond to any physical change in Bob's system at all, it is just an update of \textit{her} beliefs, hence no nonlocal action is implied. It may be that she assigns a spin eigenstate to his system, while Bob maintains the good old mixed state, but that doesn't mean that she's right and he's wrong, or \textit{vice versa}, even though they are both willing to make single-case probability claims about the results of measurement on his system that would disagree. Given the subjective Bayesian setting, they may both disagree on what the probabilities for measurement outcomes are without either one or other (or both) being wrong.

So we can see that what are arguably the most troubling conundrums typically taken to block our understanding of quantum theory are dissolved in this approach: we need worry about them no longer.

The final set of reasons for entertaining the quantum Bayesian viewpoint recur to the philosophical point of departure of the programme elaborated earlier. The hope expressed in the approach is that when the correct view is taken of certain elements of the quantum formalism (i.e., when the subjective Bayesian conception of quantum states and related structures is adopted) it will be possible to `see through' the quantum formalism to the real ontological lessons it is trying to teach us. Fuchs and Schack put it in the following way:
\begin{quote}
[O]ne...might say of quantum theory, that in those cases where it is not just Bayesian probability theory full stop, it is a theory of stimulation and response \citep{fuchs:what,fuchs:paulian}. The agent, through the process of quantum measurement stimulates the world external to himself. The world, in return, stimulates a response in the agent that is quantified by a change in his beliefs---i.e., by a change from a prior to a posterior quantum state. Somewhere in the structure of those belief changes lies quantum theory's most direct statement about what we believe of the world as it is without agents. \citep{fs:unknown04}
\end{quote}
Given the point of departure of a Bayesian view of the state, and using techniques from quantum information theory, the aim is to winnow the objective elements of quantum theory (reflecting physical facts about the world) from the subjective (to do with our reasoning). Ultimately, the hope is to show that the mathematical structure of quantum mechanics is largely forced on us, by demonstrating that it represents the only, or, perhaps, simply the most natural, framework in which intersubjective agreement and empirical success can be achieved given the recalcitrance of the world; its resistance to normal descriptive theorising. (Here we may emphasise the links to thoughts familiar from the Copenhagen tradition: we are being invited to see quantum theory as the best we can do given that---in Pauli's phrase---the ideal of a detached observer may not be obtained\footnote{This striking phrase appears in a letter from Pauli to Born \citep[p.218]{born:einstein}.}.) So the final reason to consider adopting this viewpoint is that it may lead us to new insights that would be impossible otherwise. However, it will be conceded on all fronts that the proof of \textit{this} pudding will be the eating.

\subsection{In more detail}\label{detail}

Let us now set things out a little more formally. According to the quantum Bayesian view, the world contains systems, apparatuses and agents. An agent, when choosing to apply quantum mechanics, will assign states $\rho$ (density operators) to systems, based on his or her background beliefs and knowledge; and (to repeat) different agents may come to different assignments, even when in the same situation (just as a subjective Bayesian treatment of probability would allow). Agents will, furthermore, assign POVMs $\{E_{d}\}$ (Positive Operator Valued Measures\footnote{POVMs provide the generalised notion of measurement going beyond von Neumann's projective measurements. See \citet[\S 2.2.6]{nielsen:chuang} or \citet{busch:book} for an introduction. In short, instead of associating a measurement outcome $d$ with a projector $P_{d}$, one may associate it with a \textit{positive operator} $E_{d}$ forming part of a resolution of the identity: $\sum_{d} E_{d} = \mathbf{1}$. Then one applies the trace rule as usual: the probability of obtaining outcome $d$ on measurement is given by $\mathrm{Tr}(\rho E_{d})$; $\sum_{d}\mathrm{Tr}(\rho E_{d}) = 1$. An operator $A$ acting on a Hilbert space $\mathcal{H}$ is positive iff $\forall \ket{\psi}{} \in \mathcal{H}, \bra{}{\psi}A\ket{\psi}{}\geq 0$. If $\mathcal{H}$ is complex then it follows that $A$ must be Hermitian.})  to apparatuses, corresponding to what they think a given piece of apparatus (having a range of possible outcomes or `pointer positions' labelled by the set $\{d\}$) measures; and they will, finally, associate quantum operations $\mathcal{E}$, representing time evolutions, to various physical processes\footnote{These operations $\mathcal{E}$ are trace non-increasing completely positive maps. Completely positive (CP) maps provide a very general notion of time evolution in quantum mechanics, including as special cases unitary evolution and measurement (both selective and non-selective). They are linear operators mapping density operators to density operators (perhaps up to normalisation). A map $\Lambda$ is positive if it maps positive operators to positive operators and completely positive if, when we consider adding a further system to the one under study, the extended map $\mathbf{1}\otimes \Lambda$ acting on the enlarged Hilbert space still maps positive operators to positive operators. Trace decreasing CP maps correspond to selective measurement procedures; while trace preserving CP maps correspond to evolution in the absence of selective measurement. The general form of a quantum operation is given by $\mathcal{E}(\rho)= \sum_{k} O_{k}\rho O_{k}^{\dag}$, where the $O_{k}$ are linear operators on the Hilbert space of the system for which $\sum_{k}O_{k}^{\dag}O_{k} \leq \mathbf{1}$. If one is measuring a POVM and obtains the outcome corresponing to the operator $E_{d}$, then the associated quantum operation is any one (of the large class) for which $\sum_{k}O_{k}^{\dag}O_{k} = E_{d}$; the particular operation depending on the details of the measurement. See \citet[Chpt. 8]{nielsen:chuang} for a very clear and accessible discussion of quantum operations.}. 

As we know, the quantum Bayesian deems the state subjective; but a little thought reveals that things can't stop there \citep{fuchs:only,fuchs:what,cfs:certainty}. The POVM assigned to an apparatus and the associated quantum operations must be a subjective matter too: this is required for consistency. Simply consider a preparation procedure. We commonly suppose that we can design devices in the lab that reliably produce certain quantum states. As a familiar model, consider an ordinary projective (von Neumann) measurement. Take the POVM associated with the measuring device in this case to be a set of one-dimensional projectors $\{P_{d}\}$ (so this in fact will be a PVM: a Projection Valued Measure). The state change rule (the quantum operation, including re-normalisation) associated with obtaining the outcome $d$ of this particular kind of measurement is
\begin{equation}\label{projective measurement}
\rho \mapsto \rho^{\prime}= \frac{P_{d}\rho P_{d}}{\mathrm{Tr}(\rho P_{d})} = P_{d}, \forall \rho.  
\end{equation}
(This mapping is often known as the L\"{u}ders rule.) Now if the POVM (PVM) $\{P_{d}\}$ associated with the measuring apparatus were an \textit{objective} feature of that apparatus, determined by the physical facts pertaining to it, then so would the state change rule be; \textit{and so too would be the post-measurement state}. It would no longer be a subjective matter what the state of the system is: it would be a matter of right or wrong determined by the physical facts. Post-measurement, the system would definitely be in the state $P_d$. This reasoning generalises to non-projective measurements and to the more general associated state-updates (\citet[\S 7]{fuchs:only}, \citet[\S\S 3-4]{cfs:certainty}). Here too objective constraints on what the right states are would inevitably be introduced. The general conclusion is that the POVM associated with a measuring device must be a subjective matter, in order to maintain the subjectivity of the state; as must the quantum operations assigned to measurement processes and to state-preparation processes in general. (Of course, this does not mean that whether or not an apparatus provides a particular read-out, or any read-out at all, is a subjective matter; on measurement one of the outcomes $d$ will, objectively, obtain. But what will be allowed to be subjective is what POVM element is associated with that outcome---what it represents the measurement of---$E_{d}$ or some $E_{d}^{\prime}$?)

Once this is conceded it seems overwhelmingly natural simply to take \textit{all} quantum operations, not just those associated with measurement and state preparation, as subjective, even though it is not strictly required by the consistency argument\footnote{Fuchs \citep[\S 7]{fuchs:only} argues for this conclusion and Leifer \citep{leifer:conda,leifer:condb}) provides supporting considerations.}. After all, if the state at time $t$ is just a subjective probability assignment for the outcomes of measurement at time $t$, and the state at a later time $t^{\prime}$ is just a subjective assignment for measurements at $t^{\prime}$, then the relationship between them looks very much like an updating of subjective probabilities and hence analogous to conditional probabilities in a standard Bayesian setting, which, of course, would immediately be granted as subjective. Against this one might be inclined to argue that at least \textit{some} time evolutions should be objective, for example, if one were convinced that at least the Hamiltonian that governs a particular system (or group of systems) must be an objective matter. We do, after all, often spend a good deal of time---and attach considerable importance to---calculating what the ground states or spectra of various Hamiltonians are; and this practice might seem to be undermined if we were to grant that Hamiltonians couldn't be objective features either. But if it is already conceded that quantum states are subjective, it's not clear what force this consideration can have. If it is not an objective matter whether any system ever \textit{is} in its ground state or not, for example, then it is unclear that there would be any further loss of explanatory power if it were conceded in addition that it is not an objective matter what the ground state of any system actually is. So it seems reasonable to conclude that the quantum Bayesian should adopt the subjectivity of quantum operations wholesale. 

\subsubsection{Coming to agreement: The quantum de Finetti representation}\label{deF} 

At this juncture we should air an obvious objection to all this; an objection which might have been felt increasingly urgently as we have progressed. That is: One might be prepared to grant that the quantum Bayesian position outlined so far is a logically consistent way of thinking about states and operations, but isn't it just \textit{de facto} false? Isn't it just the case that different scientists \textit{do} agree on what the quantum states prepared in the lab are, what POVMs measuring devices actually measure, and what time evolutions are? In quantum mechanical practice there's surely none of this differing (dithering?) of subjective judgements to be found. If there ever are any disagreements we just do further measurements and that settles the matter, full stop.

It should be clear that this objection is just as much, or as little, of an objection as the corresponding complaint against ordinary subjective Bayesianism. Traditionally, subjectivists have responded by proving various theorems to the effect that subjectivist \textit{surrogates} for objectivity may be found (e.g. \citet{deF:foresight}): explanations of why different agents's degrees of belief may be expected to come into alignment given enough data, in suitable circumstances; hence the \textit{appearance} (but only the appearance) of objective properties. Impressively, this same kind of reasoning proves to be available in the quantum case too \citep{cfs:deF,sbc:2001,fs:unknown04}.

An important example is given by the estimation using frequency data of what are often thought of as the \textit{unknown} chances associated with repeated independent, identically distributed (i.i.d) trials. If a number of parties consider the frequency data from many repeated throws of a biased die, for example, then we intuitively think that they will all home in on the correct probability distribution entailed by the bias, given enough of the i.i.d trials. Cases like this provide a lot of the intuitive support for objectivism about probabilities: there must be objective probability as there is something out there that we are all finding out about. But de Finetti famously provided a counter to this line of thought.  

Begin with the recognition that `an unknown probability' is a nonsense from the subjectivist view. Probabilities are people's degrees of belief, so whenever there is a probability, there had better be somebody to bear it, hence it can't be unknown. Thus to begin with it seems impossible even to state what is going on in the estimation scenario just described: so much the worse for the subjectivist, it seems. But de Finetti showed how to get around this problem. Consider the event space composed of all the different possible sequences of outcomes for a large number $n$ of repeated trials of the experiment. The objectivist will think that there is a  probability distribution over these sequences generated from the i.i.d chances associated with each of the trials; the subjectivist will not, instead assigning a certain prior probability distribution over the sequences. de Finetti identified what the relevant aspect of this prior will be in the estimation scenario. If one \textit{judges} (no more) that each trial is relevantly similar, then the prior probability distribution one assigns will be permutation symmetric: it would make no difference to your probability assignment if you were to imagine two or more of the outcomes in a sequence being swapped in order. If we further judge that an \textit{arbitrary number} of the trials would be relevantly similar (again capturing a natural aspect of the setting) then the prior will in addition be what is termed \textit{exchangeable}; that is, both symmetric and derivable by marginalising a symmetric prior over a larger number $n + m$, ($m > 0$) of events. Finally, if the prior is exchangeable, then de Finetti showed in his representation theorem that it may be written uniquely \textit{as if} it were an ignorance probability over a range of unknown i.i.d. chances. So if we imagine that some variables $x_{i}$ are the random variables representing the outcome of each individual trial and $p(x_{1},x_{2},\ldots, x_{n})$ were our exchangeable prior, capturing our thought that the various trials were relevantly similar, then the prior may be expressed as:
\begin{equation}\label{de F representation}
p(x_{1},x_{2},\ldots,x_{n}) = \int P(\mathbf{p})p_{x_{1}}p_{x_{2}}\ldots p_{x_{n}}d\mathbf{p},
\end{equation}
where the $p_{j}$ are what one would think of as the chances of getting outcome $j$ on a particular run of the experiment, were one an objectivist, and $P(\mathbf{p})$ is a distribution (normalised to one) over the space of all the possible chance distributions $\mathbf{p}=\{p_{1},p_{2},\ldots\}$, $0\leq p_{j} \leq 1$, $\sum_{j}p_{j} = 1$. Then one would appeal to Bayes' theorem to show how by conditioning the exchangeable prior on the frequency data received, the weighting $P(\mathbf{p})$ becomes increasingly peaked as data is received; and that different agents beginning with different exchangeable priors will find their respective distributions $P(\mathbf{p})$ peaking around the same point\footnote{More detailed statements of the results involve noting that the various agents must be suitably reasonable to begin with, for example, as Fuchs puts it \citep{fuchs:only}, being willing to learn, which would manifest itself in the requirement that the initial distributions $P(\mathbf{p})$ may be arbitrarily close to zero at any point but must at none actually be zero.}. 

Thus the Bayesian account of the appearance of objectivity: homing in is coming to agreement.

Caves, Fuchs and Schack apply the same conceptual manouevres to the quantum case, for the same ends\footnote{Quantum versions of the de Finetti theorem had been proven before by a number of authors (for refs. see \citet{cfs:deF}), but Caves, Fuchs and Schack were the first to deploy the ideas as part of the debate over the nature of quantum states; and they provided a simplified proof of the theorem.}. The quantum analogue of the classical estimation case just described is what is often known as \textit{quantum state tomography}. It is well known that it is impossible to determine the unknown state of an individual quantum system; this fact lies at the heart of quantum cryptography, for example, and indeed lies behind many of the significant differences between quantum and classical information\footnote{For a philosophers' introduction to quantum cryptography and related ideas, see \citet[\S 2]{timpson:paqit}.}. However, if one is presented with a sufficient number of identically prepared systems then it \textit{is} possible to identify the state by taking suitable measurements. For example if one has a large number of $k$ dimensional systems, all prepared in the same way, then determining the expectation values of $k^{2}$ linearly independent operators by measurement will suffice to determine the state \citep{fano,band:park}. It is in this kind of way that one typically establishes that one's preparation device in the lab is actually doing what one hopes.

The quantum Bayesian response is first to present a quantum version of de Finetti's representation theorem. If one begins with an exchangeable\footnote{`Exchangeable' for density operators means just the same as for probability distributions: the operator is symmetric under interchange of systems and may be derived by taking the partial trace over a symmetric assignment to a larger number of systems.} density operator assignment $\rho^{(n)}$ for $n$ systems which one \textit{judges} (not knows) to have been identically prepared, then it may be shown that the state may be written uniquely in the form:
\begin{equation}
\rho^{(n)} = \int P(\rho)\rho\otimes\rho\otimes\ldots\otimes\rho\, d\rho,
\end{equation}
where $P(\rho)$ is a distribution (again, normalised to one) over the space of density operators. The second step is the proof that a suitable analogue of the Bayes rule reasoning applies \citep{sbc:2001}: given suitable data from measurements on the prepared systems, $P(\rho)$ will approach a delta function, the same even for different priors. Related results \citep{fss:process, fs:unknown04} may also be proven for \textit{quantum process tomography}, the business of establishing what evolution is applied by a given procedure in the lab, again defending the subjectivist view, this time for the subjectivity of operations.

These are impressive results for the quantum Bayesian programme; and the emphasis which has been laid on quantum analogues of the de Finetti theorem by this approach has already borne independent theoretical fruit in providing improved means of establishing the difficult question of whether particular quantum cryptographic protocols are really secure against the most general kinds of quantum attacks \citep{renner}. But it is worth noting that the quantum subjectivist responses to the objectivist challenge do of course share a common weakness with the usual classical subjectivist stories. That is, the objectivist might be quick to respond that all that has been shown in these theorems is how it might be \textit{possible} for certain agents in idealised situations to come to agreement\footnote{In particular, one might be concerned about the rate at which agreement would be reached for differing priors and the intial constraints that the various agents be `reasonable' in certain technically defined ways.}, not that \textit{actual} agreement in the real world would ever be met; and they might add that, moreover, Bayesianism is plausible only a as \textit{normative} theory, specifying how agents should act if they are to satisfy certain consistency requirements; it is not a descriptive theory stating how agents will act (nothing \textit{forces} someone to reason Bayesianly after all), so there is an important sense in which it \textit{cannot} provide the kind of explanation for agreement that might be required. The subjectivist's (quite reasonable) \textit{tu quoque} might be that objectivist theories of statistical inference are scarcely in good shape (c.f. \citet{howson:urbach}); however I shall leave this issue to one side as it is part of the general debate between subjectivist and objectivist notions of probability and not specific to the quantum case. What we should not lose sight of is the impressive result which has been established that appealing to the laws of quantum mechanics does not settle the issue of the existence of objective probabilities.

\subsubsection{What gets to be an apparatus?}\label{apparatus}

We have said that an agent will assign states to systems and associate POVMs to measuring apparatuses, but what gets to be an apparatus? What gets to be a system? This is one of the common forms in which our disquiet about dealing with the quantum world has manifest itself. 
Bohr, for instance, is notorious for his free-wheeling attitude to where one might draw the line between quantum and classical \citep{bohr:epr,bohr:einstein}; and his apparently shifty and unprincipled (in both senses!) sliding of the split has often induced nausea in the realist minded. ``One should not play so fast and loose with reality!" runs the thought. 

Others in the Copenhagen tradition have argued in Bohr's defence, however. Peres makes what would seem to be the appropriate rejoinder quite crisply:
\begin{quote}
Bohr never claimed that different physical laws applied to microscopic and macroscopic systems. He only insisted on the necessity of using \textit{different modes of description} for the two classes of objects. \citep[p.11]{peres:book}
\end{quote}
This response doesn't solve all difficulties (as Peres himself immediately goes on to note), but it is a move in the right direction if one were inclined to defend Bohr (I am not); and it signals the right kind of direction for the quantum Bayesian to take too. The best way of thinking in their approach is that there is a \textit{single} world, not a divide between quantum and classical worlds; and this single world is a \textit{quantum} world, that is, a world in which we have learnt that we must use quantum mechanics as a largely pragmatic tool if we wish to make very detailed predictions; a world which precludes descriptive theory below a certain level. Above that level there will be a mass of stateable truths, many of which will be pretty well subsumable under theory; yet there is only one world. A world which contains big things and small things and where the big things are composed of the smaller ones. 

Thus the rule will be that one treats as a system that which one needs to apply quantum mechanics to in order to ensure best predictive success. For the truly microscopic we have learnt that no classical description gets things even vaguely right and hence the microscopic must always be treated as system. For larger objects it may be that fairly good, or even very good, descriptions are to be had, so one may often not need to apply quantum mechanics to them, not need to treat them as system; but one may always choose to do so if one wishes (although typically it will make no predictive difference); and sometimes one may need to do so, for example if one is interested in particular fine detail of a micro-macro interaction. 

So: what get to be apparatuses rather than systems are items i) about which there are stateable facts at all times (so one always has the option of \textit{not} treating them as quantum) and ii) which one happens to be using to probe the objects being treated as systems, or which one deems apt for such a purpose.

As already mentioned, one is always free to step back and apply quantum mechanics to the apparatus too; one will then have other objects in mind as the apparatuses with which one might probe this large system and its relation to other systems; but it is important to recognise that this shift doesn't change any of the facts about the object concerned, nor indicate any change in our attitude towards the facts about it (so no call for queasiness!). We may still believe that the object previously treated as a measuring device, but now being treated as a system to be assigned a quantum state, has stateable truths about where its pointer may be pointing (for example). And that's so even if I assign it a quantum state; and even if I end up assigning it a state which is not an eigenstate of the pointer observable.

Why so? Well, attend to the discipline that states only represent degrees of belief about what the outcomes of future measurements would be. If I believe that the pointer is at position $x$, and I wish to ascribe a quantum state to the apparatus, then, given one's normal assumptions about the functioning of one's perceptual faculties, I certainly will assign it the corresponding eigenstate of the pointer observable: I believe it has a particular position and I believe I will see the pointer in just that position when I look (barring silly accidents); a certain state assignment follows. But the converse kind of reasoning doesn't hold: the fact that I do not assign a pointer eigenstate does not mean that I don't believe that the pointer has some definite position or other: that would hold only if one subscribed to the eigenstate-eigenvalue link, which is out of the question in this approach. The state I assign only corresponds to my probabilistic predictions for where the pointer will be found when I look; and the theory which informs my judgements---quantum theory---might dictate that a superposition of pointer eigenstates, or more likely, an entangled state involving such a superposition, is predictively the best. So to recap, contrary to one's intuition shaped by the common use of the eigenstate-eigenvalue link, I can believe that there is some fact (of which I am ignorant) about where the pointer of a device is pointing, without having to assign it a pointer eigenstate, or even a convex combination of projectors onto such eigenstates. When I look I will certainly find it in a definite position (there is a fact about where it is at all times, after all) but my probability distribution over the possibilities may be a non-classical one given by some state assignment involving a superposition. The facts about the pointer position don't determine a state assignment, nor does any belief that there are some such facts; only my beliefs looking forward to measurement do.    

Thus the quantum Bayesian is in a pretty good position when it comes to the system/apparatus divide: in this picture a shift in the status of an object from apparatus to system should have no tendency to induce nausea as it has no association with a downgrading of definite, observable properties to wavy indistinctness. Furthermore, the shiftiness is \textit{principled} given that one begins from a setting in which quantum mechanics is to be treated as a pragmatic means of dealing with the world rather than being a candidate descriptive theory: one only treats an apparatus quantum mechanically when one needs the greater predictive detail; but that's not being mercenary---it is just the name of the game. And there is a final advantage. One of the things which is a trouble for approaches such as Peres' which also allow a shift from treating apparatus as apparatus in a measurement to treating it as system, is in obtaining consistency between the two descriptions provided of what is, after all, only one physical measurement process. In particular, one will obtain different system-apparatus correlations depending on whether one treats the apparatus as quantum or not; that is, joint probability distributions for measurements on system and apparatus together will be different in the two cases. Now one would typically appeal to decoherence to suggest that the difference isn't ever going to be much in practice, but the position is still rather unsatisfactory. But happily, the problem does not arise for the quantum Bayesian. Given that probability distributions are subjective, the different levels of description with their different joint probabilities do not disagree about how anything is in the world.

We normally tend to think that our fundamental physical theories are, or ought to be, universal in scope: that they apply equally to \textit{all} physical objects in any situation and provide a framework for a comprehensive treatment of the physical world at a basic level\footnote{This presumption has not, however, gone unchallenged. For stimulating arguments against see \citet{dupre:disunity} and especially \citet{cartwright:hlpl,cartwright:dappled}; while \citet{hoefer:fundamentalism} and \citet{sklar:uniform}, for example, defend the orthodoxy.}. It is clear that in one sense the quantum Bayesian picture does not satisfy this ideal. In order for the theory to be intelligible, we must draw distinctions between system and apparatus, treat these items differently, and highlight the role of agents as the owners and appliers of quantum states. Quantum mechanics on this picture cannot be a universal theory in the sense that it can be applied to \textit{everything at once}; some things are not assigned a quantum state in order that `assigning a quantum state' can be a meaningful phrase. The theory doesn't proffer a view from nowhere, in Nagel's phrase \citep{nagel:view}. However, there remains a perfectly good sense in which the theory thus construed \textit{is} universal, namely, it is applicable to \textit{anything and everything}, only piecewise, not all at once\footnote{One can even make sense of assigning a wavefunction to the universe, so long as one does not take it to include all degrees of freedom, cf. \citet{fuchs:peres}.}. Maybe that should be good enough for us. Certainly, it will allow us to make all the predictions we could ever want.       

\subsection{From information to belief}\label{itob}

The earlier expressions of the quantum Bayesian viewpoint \citep{fuchs:light,fuchs:paulian,cfs:bayesprobs} suffered from an uncomfortable problem. In these, the notion of information played a more central r\^{o}le; indeed the central statement of the position was that the quantum state represented \textit{information} about what the results of future experiments would be. This proved to be a false start, as the concept of information is not apt to play the required r\^{o}le. As I have discussed elsewhere \citep{timpson:iii,timpson:book} the term `information' is, just like the term `knowledge', \textit{factive}. That is, one can't have the information that $p$ unless $p$ is the case; one can't know that $p$ unless $p$; one can't have information about something unless that information is correct. This factivity re-introduces the objectivity of quantum states it was the express aim of the approach to avoid and the proposed resolutions of the problems of measurement and nonlocality founder on this. Thus if one wishes to associate the quantum state with a cognitive state in order to ameliorate the conceptual difficulties of quantum mechanics, one has to choose belief rather than knowledge and one must eschew `information' as a way of expressing what one takes the state to represent. Fuchs recognised the difficulty---although not putting it in quite these terms---and it forms the main theme of the discussions in \citet{fuchs:what}.

The shift from the notion of information to that of belief also goes along with a shift from so-called \textit{objective Bayesianism} about probabilities to the full blown subjective Bayesianism we have been discussing. In an objective Bayesianism viewpoint \citep{jaynes:collection}, probabilities are degrees of belief, but it is maintained that circumstances can entail that certain assignments are correct and others wrong (e.g., one ought to assign a probability distibution that maximises entropy subject to the constraints, it might be maintained \citep{jaynes:1957}). The problem with this sort of view in the quantum case is that it would mean that some state assignments are right and others wrong, in virtue of how the world is; but of course it was just this kind of objectivity which the approach seeks to avoid in order to make progress. If some states are right and others wrong then we can no longer discount Wigner's friend concerns and the resolution of the conundrums of nonlocality in quantum mechanics becomes highly problematic. It looks as if no half-way house would be satisfactory. If one wants to adopt something like the Bayesian line in quantum mechanics to any purpose then it seems one must go the whole hog and accept the radical \textit{subjective} Bayesian line.

\section{Not solipsism; and not instrumentalism, either}\label{not solipsism}

Having explored the quantum Bayesian position in some detail, let us now consider some common objections or misconceptions to which the approach has been prey.

The first is the charge that the position boils down to a fancy form of solipsism: the view that only my mind and my mental states exist, there is no external world and there are no other minds. It would certainly be a killer blow to the position if this charge could be made to stick; but it cannot. The challenge  is suggested by the pre-eminence of the notion of the agent in the statement of what quantum mechanics concerns: states (etc.) are individual agents' degrees of belief about what they would see in various circumstances; in the limit, a quantum state is \textit{my} set of degrees of belief. Now one gets from here to the charge of solipsism by adding the assumption, quite natural in some settings, that everything is made out of matter characterised  quantum mechanically (picturesquely: `everything is made out of wavefunctions'). That is, that statements made in quantum mechanical language describe what there is in the world; and that everything is to be characterised by the quantum state assigned to it. But if to make a quantum mechanical statement attributing a state only involves a claim concerning my beliefs (quantum Bayesianism), then there can be no things but the beliefs themselves and the bearer of the beliefs (solipsism); for no other kind of content latching onto externally existing things is expressed. It would be of no avail to say that the beliefs concern what might happen to various concrete objects, some of them macroscopic, as those objects themselves will, ultimately, only be analysed in quantum mechanical terms too, the argument proceeds. Thus one ends up with no substance to the world other than a pattern of beliefs. To put it another way, if one deletes the agent ascribing states, then there are no states, so no things having any properties at all.\footnote{It's true that one might point out that it would still be allowed by this argument that there are externally existing things---those to which states might be attributed if I actually existed---it's just that they would have no properties, given my absence, as there would be no states. This isn't quite solipsism, perhaps, but it's not a great position to be in either.} 

It should be quite evident where this reasoning breaks down: the quantum Bayesian simply rejects the idea that the quantum mechanical statements one would typically make describe how things are. They don't think that everything is ultimately made from matter which is \textit{characterised}---attributed properties by---the quantum state. Moreover, the view is not a \textit{reductionist} one and because of this it is by no means solipsist.

The typical view encapsulated in the idea of \textit{theory reduction} is that the laws, statements, predictions and so on of one theory are determined by, or can be derived from, another, more fundamental theory; perhaps with a little correction along the way. And it is a very natural thought that the physical facts about the familiar objects that surround us (e.g. tables, chairs, lab benches, computers) are determined by more fundamental facts about their constituents; and these constituents being quantum systems (when you get down to it) that the facts about these familiar objects will be determined by facts stated in the terms of quantum mechanics about those constituents. 

Many views of quantum mechanics would be very happy to allow this, but it is no part of the quantum Bayesian picture at all. The point of departure of the position, after all, is that quantum mechanical statements do not provide us with a story about how things are with microscopic systems, a set of facts characterising them; hence these statements are \textit{simply not candidates for a class of statements that might serve as a reduction base for classical level (non-quantum) statements.} The issue of solipsism cannot arise. For the quantum Bayesian, microscopic objects exist mind- and agent-independently, although it is granted that there is little that can be said about them\footnote{Which need not be to say that there aren't various truths about them; but the majority of such truths are just unspeakable---beyond the capture of theory and description. We will see more of this below.}: one deals with them by assigning an agent-dependent state. Macroscopic objects also exist mind- and agent-independently; and there is plenty that can be said about \textit{them}, although the facts about them aren't supposed to be reducible, even in principle, to facts stated in the language of quantum mechanics about their microscopic constituents.    

A second common charge is that quantum Bayesianism is no more than a version of instrumentalism. Instrumentalism, recall, is the doctrine that scientific theories do not describe the world; rather they are merely instruments that one uses for making predictions about what will be observed. Thus the theoretical claims made in a theory are not apt to be judged as true or false on this conception, they are just tools that are used to help organise empirical predictions. Now it is clear that \textit{something} along these lines is true of the quantum Bayesian position, but it is less clear that what is true counts as an objection. Thus the quantum Bayesian begins by adopting a form of instrumentalism \textit{about the quantum state}, but that is far from adopting instrumentalism about quantum mechanics \textit{toute court}. The state is, of course, not to be given a realist reading, it is construed instrumentally, or pragmatically, as concerning predictions only; but that is conceded at the outset of the programme. And it is conceded in order to serve realist ends. The non-realist view of the state is not the \textit{end point} of the proposal, closing off further conceptual or philosophical enquiry about the nature of the world or the nature of quantum mechanics, rather it is the \textit{starting point}. Thus it would be misguided to attack the approach as being instrumentalist in character. There is certainly no assertion that the aim or end of science is merely prediction, that we should stop right there. Given the hypothesis that the world precludes descriptive theorising below a certain level, to turn away from literal realist readings of one's best fundamental theory is not to turn away from realism, but to seek the only kind of access one can have.

Another kind of objection invites us to reconsider Wigner's friend. The quantum Bayesian view will allow different agents to assign different states; and Wigner and friend will typically assign different states to the contents of the lab after the friend's experiment. These different states, however, correspond to different predictions for joint measurements on system and apparatus in the lab; and surely we can just test these predictions (at least in principle) to see who is right. Given that, the quantum Bayesian must be in error when they submit that Wigner and friend can disagree unproblematically: on the contrary, one is right and the other is wrong. Hagar, for example \citep{hagar:looks}, refers to this kind of objection. 

As is no doubt obvious, the objection only looks plausible if one is not working with subjective probabilities, however. With subjective Bayesian probabilities the facts don't determine or make right or wrong a probability assignment, so there is no measurement one could do which would show one assignment or the other to be wrong. This illustrates a general property of the quantum Bayesian position. No objection can be successful which takes the form: `in such and such a situation, the quantum Bayesian position will give rise to, or will allow as a possibility, a state assignment which can be shown not to fit the facts,' simply because the position denies that the requisite kinds of relations between physical facts and probability assignments hold. (No more, obviously, could it fall to the converse argument that the facts in such and such a situation require a particular probability distribution, yet quantum Bayesianism allows another.) The subjectivist conception of probabilities is a consistent one. If one wants to convict the quantum Bayesian of error then one will need to furnish a general argument for the failure of the subjectivist conception of probability in addition; and this is no straightforward task.

In his discussion, Hagar notes the crucial point that the relevant joint probabilities on which the challenge turns are supposed to be subjective, so not subject to decision by experiment, but still seems to think that there is a difficulty in making quantum mechanical probabilities relative to the agent in this setting, as this will introduce an arbitrary cut between the observer and nature which will imply that `what counts as real, i.e., as having definite properties is now \textit{dependent} on where this cut is made' \citep[p.767]{hagar:looks}. And this seems too far to go for consistency even with the quantum Bayesian's subtle realism. But it is unclear why it should be supposed that where one decides to make the cut, that is, where one chooses in a given situation to draw the line between system and apparatus, should have any ontological implications at all. Just to treat an object as quantum mechanical---that is, to assign it a quantum state---is to take no stance at all towards its ontology, in this setting. It is merely to apply a certain structure of probabilistic reasoning to consideration of its interaction with devices apt to investigate it. As we saw above in detail, the quantum Bayesian has no difficulties with a shifting quantum/classical split; and to attribute a (perhaps entangled) quantum state to a macroscopic device is consistent with continued belief in that device's definite classical level properties. There are no troubles for the quantum Bayesian here.

\subsection{Summary: The virtues}\label{virtues}

As a framework for thinking about and investigating quantum mechanics, quantum Bayesianism has considerable virtues. In many ways it represents the \textit{acme} of certain traditional lines of thought about quantum mechanics. It leaves us with a radical picture, but that is salutary in indicating what is involved in consistently developing those ideas in a useful way. Thus if one were inclined to Copenhagen-flavoured analyses of the quantum state in terms of some cognitive state, in order to avoid difficulties that realist conceptions of the state entail (e.g., \citep{hartle:1968,mermin:whose,peierls:defence,wheeler,foundationalprinciple}), then quantum Bayesianism, where the cognitive state called on is belief, not knowledge, is the only consistent way to do that\footnote{The only consistent way, that is, if one wishes either to allow single-case probabilities or to avoid subscribing to an unilluminating instrumentalism.}. It shows that there is no cheap resolution of our traditional troubles to be had here. Or perhaps one is drawn to the Paulian thought that quantum mechanics reveals us to be living in a world where the observer may not be detached from the phenomena he or she helps bring about. The quantum Bayesian programme seeks to move such reflections from the realm of loose metaphor to the realm of concrete and useful theoretical statements; and it sets out a precise formal starting point and a recommended direction for doing so.

This point bears double emphasis. A main, perhaps \textit{the} main, attraction of the approach is that it aims to fill-in a yawning gap associated with many views that can be grouped broadly within the Copenhagen tradition: It is all very well, perhaps, (one may grant) adopting some non-realist view of the quantum formalism (as Bohr and many others have urged); but, one may ask, (with increasing frustration!) why is it that our best theory of the very small takes such a form that it needs to be interpreted in this manner? Why are we forced to a theory that does not have a straightforward realist interpretation? Why is \textit{this} the best we can do? The programme of Caves, Fuchs and Schack sets out its stall to make progress with these questions, hoping to arrive at some simple physical statements which capture rigorously\footnote{We'd like a better answer than: the finite size of the quantum of action entails a disturbance not to be discounted \citep{bohr:como}, for example!} what it is about the world that forces us to a theory with the structure of quantum mechanics. And in doing so it invites us to consider interesting new kinds of question in the foundations of quantum mechanics. We have already seen, for example: `How ought one to make sense of the notion of an unknown state in quantum mechanics if states are subjective?', `How exactly does the region of quantum allowed probability judgements compare to the classically allowed; and why?', `What Bayesian reason might there be for complex Hilbert spaces, as opposed to real or quaternionic?'.    

While the aim of the programme is to seek a transparent conceptual basis for quantum mechanics, with the help of techniques from quantum information theory and with the suspicion that some information-theoretic principles might have an important r\^{o}le to play, it should be noted that there is no claim that quantum mechanics should be understood as a Principle theory, unlike Clifton, Bub and Halvorson's contention \citep{cbh}. The notion of a Principle theory is due to Einstein, who emphasised a distinction between Principle and Constructive theories when describing his 1905 methodology for arriving at Special Relativity. Principle theories begin from some general well-grounded phenomenological principles in order to derive constraints that any processes in a given domain have to satisfy; Constructive theories build from the bottom up, from what are considered to be suitably basic (and simple) elements and the laws governing their behaviour, to more complex phenomena. Thermodynamics is the paradigm Principle theory and kinetic theory the contrasting Constructive theory of the same domain\footnote{Einstein's main principles for Special Relativity in 1905 were, of course, the Principle of Relativity and the Light Postulate; amongst the entailed constraints on the domain of phenomena, time dilation and length contraction.}. Following their proof that (the kinematic structure of) quantum theory may be axiomatised (in a general $C^{*}$-algebraic setting) using three information-theoretic principles, Clifton, Bub and Halvorson went on to suggest that the theory should be understood as a \textit{Principle theory concerned with the behaviour of information}, rather than a Constructive theory postulating a conceptually troublesome mechanics of particles and fields (the more familiar view). This suggestion is problematic (see \citet[\S 9.2]{timpson:thesis}, \citet{timpson:book} for discussion) but the point is that it is not a view shared by the quantum Bayesian programme. Not all attempts to gain a better understanding of quantum mechanics by appeal to information-theoretic principles that help characterise the theory need conceive of the theory as a Principle theory. Finally, in further contrast to the approach of Clifton, Bub and Halvorson, rather than seeking to provide an axiomatisation of the quantum formalism which, when it is recovered, will perforce be open to interpretation in various ways\footnote{You've got the structure of Hilbert spaces, states, observables, operations etc. back. Now, how does it all relate to reality? Realistically? Instrumentally? Bohr? Everett? Bohm?}, the quantum Bayesian instead takes one interpretive stance to begin with and then proposes to see whether or not it will lead us to a perspicuous axiomatisation. This is obviously quite another approach.

\section{Challenges}\label{challenges}

We have seen that the quantum Bayesian position has many benefits; that it is an important and an intriguing proposal. It is time now to consider some significant challenges the position faces. The first (Section~\ref{ontology}) concerns whether or not we can make out some kind of reasonable ontology for the theory; whether we can render sufficiently intelligible the proposed world-picture in which the fundamental level of physical reality is unspeakable in some manner; is resistant to the capture of law and to descriptive theorising. And if we are able to make out such an ontology, is it one which we could sensibly adopt? Next we shall consider (Section~\ref{explanation}) whether the quantum Bayesian conception would leave us with sufficient explanatory resources in quantum mechanics. Our final set of concerns (Section~\ref{subprobs}) turn on whether the subjective Bayesian conception of probabilities is really, in the end, acceptable within quantum mechanics.

\subsection{What's the ontology?}\label{ontology}

In fact it seems that a reasonably sensible, indeed, an almost off-the-shelf, ontology is available to the quantum Bayesian, but first a caveat. We already noted in the introduction that the exposition of the quantum Bayesian position so far is most closely guided by Fuchs' writings, while still aiming to represent the views of the other main proponents of the position, Caves and Schack, but perhaps to a lesser degree. That warning should be doubly emphasised when we turn to a detailed consideration of what ontology, or ontologies, might be available to underpin the quantum Bayesian picture, for the textual sources here are almost exclusively Fuchs' writings in correspondence with Caves, Schack and other colleagues \citep{fuchs:paulian,fuchs:what,fuchs:cg}. But while it is Fuchs who has perhaps expended the greatest effort on trying to clarify the kind of ontology that might go along most naturally with the quantum Bayesian position, I fear he might not favour the kind of ontology that I shall offer; for it will be rather more conservative than the sort of position I take it he would most prefer.

What furniture does the quantum Bayesian need? Let us start with the basic thought that at least we must have systems and measuring devices. The systems are assigned quantum states and the measuring devices quantum operations and POVMS. Systems and devices interact and events will occur. So we have systems, devices (or apparatuses) and we have events. Now naturally we should take the measuring devices (and all the other larger kinds of objects that we admit, including ourselves) to be made from various of the systems: the quantum Bayesian does not go so far as to deny this humdrum truth. But the by-now familiar thought is that the behaviour of the systems falls under no law and they do not properly admit of direct description.

In particular, the really crucial conception is that what (singular) event will occur when a system and a measuring device interact is \textit{not determined by anything}, not even probabilistically. That is, there are no facts about the world, prior to the measurement outcome actually obtaining, which determine what that outcome would be, or even provide a probability distribution over different possible outcomes. This feature strongly emphasises the departure from standard ways of thinking about quantum theory: for example, hidden variable theories of any flavour would precisely be in the business of providing facts of the kind that are being denied here: either facts that determine what the outcome of a measurement would be (deterministic hidden variable theories), or do so given the additional specification of a context (contextual hidden variable theories), or at least provide a probability for what the outcomes will be (stochastic hidden variable theories, within which we may include realist collapse theories such as GRW). But none of these kinds of facts is supposed to exist on the quantum Bayesian picture.      

Fuchs is tempted to draw from this last crucial insistence on the absence of any determination of what event would---or might be likely to---occur on measurement, philosophical conclusions of a pragmatist\footnote{Pragmatism is the position traditionally associated with the nineteenth and early twentieth century American philosophers Pierce, James and Dewey; its defining characteristic being the rejection of \textit{correspondence} notions of truth in which truths are supposed to mirror an independently existing reality after which we happen to seek, in favour of the thought that truth may not be separated from the process of enquiry itself. The caricature slogan for the pragmatist's replacement notion (definition) of truth is that `Truth is what works!' in the business of the sincere and open investigation of nature.}  and open-future (or `growing block theory') variety\footnote{The philosophical doctrine of a growing block view of reality is the idea that the present moment and the past exist, but future events and objects do not; and there are no facts about what will obtain in the future: it is open. As time passes, the `line' of the present rolls forward and more events are thereby incorporated into the real. The view contrasts with \textit{presentism} the idea that only the present exists and \textit{eternalism}, the view that all events in the 4-dimensional `block universe' are all on a par, ontologically.}. Thus:\footnote{In what follows, recipient and date locate where quoted items may be found in \citet{fuchs:cg}.}
\begin{quotation}

\noindent Something new really does come into the world when two bits of it [system and apparatus] are united. We capture the idea that something new really arises by saying that physical law cannot go there---that the individual outcome of a quantum measurement is random and lawless. (To Caves-Schack 4.9.01)\\

\noindent A quantum world..[is] a world in continual creation (\citet{fuchs:dq} p.1)\\

\noindent There is no such thing as THE universe in any completed and waiting-to-be-discovered sense...the universe as a whole is still under construction...Nothing is completed...even the ``very laws" of physics. The idea is that they too are building up in precisely the way---and ever in the same danger of falling down as---individual organic species. (To Wiseman 24.6.02)\\

\noindent My point of departure, unlike [William] James's, [is] not abstract philosophy. It [is] simply trying to make sense of quantum mechanics, where the most reasonable and simplest conclusion one can draw from the Kochen-Specker results and Bell inequality violations is...``unperformed measurements have no outcomes". The measurement provokes the truth-value into existence; it doesn't exist beforehand. (To Wiseman 27.6.02)\\

\noindent How does the theory tell us there is much more to the world than it can say? It tells us that \textit{facts} can be made to come into existence, and not just some time in the remote past called the ``big bang" but here and now, all the time, whenever an observer sets out to perform...a quantum measurement...[I]t hints that facts are being created all the time all around us. (To Musser 7.7.04)

\end{quotation}

But adhering to the thought that unperformed measurements have no outcomes and that the actual results of measurements that obtain are not determined by anything---are `new life' events---does not require one either to be pragmatist along the lines of James, Pierce or Dewey, or to deny eternalism. So even if the advertised view of quantum measurement is correct, it does not support the pragmatist position or the idea of the growing block, as a less philosophically contentious alternative is available.

It would seem that the cleanest setting for the proposal is just this. Grant facts to be timelessly true as the eternalist would assert: facts may pertain to particular times, or to objects existing at particular times, or to happenings at particular times (that is to say, there are various truths about these things) but the facts themselves are not temporally qualified: statements about the world are just true or false simpliciter; facts don't come into or go out of existence at any time (they are not themselves part of the spatio-temporal order) they just \textit{are} thus-and-so. For any time let us suppose that there is (timelessly) a fact about how things are at that time; there is still a lot of freedom about how the facts about different times might relate to one another. Thus consider a particular time $t$ at which some quantum measurement outcome has occurred. There is a fact pertaining to $t$ about what that outcome is. All we require to capture the quantum Bayesian position is simply to assert that there are no facts pertaining to any previous times that \textit{determine} what that fact about the measurement outcome is. And this assertion is to be generalised: for no time $t$ do any of the facts about previous times (later ones too, one might want to add) determine the \textit{facts at} $t$, nor do any of these previous (or subsequent) facts even confer a probability on how things are at $t$. Hence the claim of lawlessness: we can picture the world as involving at the fundamental micro-level a four-dimensional pattern of events, where the events at a given time do not determine, or imply probabilities for, events at any earlier or later time. The four-dimensional pattern is too unruly: it does not admit of any parsing into laws, or even weaker forms of generalisation, not even statistical ones.

This, then, is the micro-level we have dubbed unspeakable; to which we are denied direct descriptive access. The picture is of a roiling mess. Fuchs adds:
\begin{quote}
For my own part, I imagine the world as a seething orgy of creation...There is no one way the world is because the world is still in creation, still being hammered out. It is still in birth and always will be...(To Sudbery-Barnum 18.8.03)
\end{quote}
But how to think of this unspeakable micro-level? It is here that we can begin to engage with some off-the-shelf ontology; for an immediate thought that might be suggested is that we should opt for a conception in which the basic systems have largely modal or dispositional characteristics, rather than occurrent, categorical ones. Thus the systems primarily have \textit{dispositions} to give rise to various events when they interact with one another. Some of these events produced will be the outcomes of interaction of smaller systems with the larger composite systems (for example, measuring devices) with which we are familiar; and it is these events that the Bayesian agent will update upon. However, it is important that the dispositions, or powers, possessed by the systems will not stably give rise to repeatable, regular behaviour when the systems interact with one another; the rules of composition of the powers are too loose (or are non-existent) we may imagine, giving rise to the lawless pattern of events. 

This idea of the systems primarily possessing modal properties---powers or dispositions---fits well with Fuchs' imagery of creation and birth, for the obtaining of events (their `birth') will be the \textit{manifestation}, or joint manifestation, of these powers of the systems; while the modal properties possessed by systems at a given time will perhaps imply a certain range of \textit{possibilities} for future times---only one of which could come to be realised---even if we concede that their joint interaction does not go so far as conferring probabilities on future events. Thus we can think of the world as at each time \textit{pregnant with possibilities}; while yet, what will happen is completely undetermined.   

The ontological picture being borrowed from here is of course that of Nancy Cartwright \citep{cartwright:dappled} who advocates an ontological picture for science in which objects primarily have dispositions or powers and it is only when these powers interact in highly contrived, or highly specialised, situations that they will give rise to the repeatable, regular behaviour that can be described by the kinds of general statements we traditionally think of as laws of nature, or as lawlike truths. Where things differ in our case is that we are imagining that at the fundamental level\footnote{Cartwright, we should note, might well be sceptical of loose talk of a `fundamental level' at all, as she wishes to combat the `fundamentalist's' notion that the laws of nature form something like a pyramid, with the basic laws of physics at the bottom, and with the laws, general claims and theoretical statements of the more specialised sciences sitting on top, supported by this foundation. But this is perhaps another locus where what I am advocating for the quantum Bayesian differs from Cartwright's picture.} there are \textit{no} situations, however specialised, in which we will obtain lawlike behaviour. Interactions of the powers of our micro systems \textit{always} give rise to unruly results.    

Why is this micro-level unspeakable? Well, most importantly, by \textit{fiat}; this is just the hypothesis we are exploring: no laws can be found to govern the behaviour of the micro-level---no such laws exist; nor can we gain any descriptions of the properties or behaviour of the micro-systems adequate for any kind of systematic theorising. But from what has been said so far, we can elaborate a little more on why the unspeakability. If the properties possessed by the micro-systems are primarily dispositional on this picture, then there is just not much to be said about how things are occurrently. The micro objects are seats of causal powers and there is nothing to explain (we are supposing) why they give rise to the particular manifestations of their powers that they do, when they do\footnote{It might be that in addition to their modal properties, the systems possess some underlying occurrent, categorical properties of some sort; but by hypothesis, these do not `ground' the modal properties in the sense of explaining how and why the powers of the systems obtain; how and why the powers will manifest themselves in the particular way they do, when they do.}. Moreover, dispositions and powers are primarily identified by what they are dispositions and powers to do; and if we cannot find anything systematic to say about how these modal properties will manifest themselves, then it is unsurprising that we will lack a grasp of what these properties themselves are. In general, the uniform absence of lawlike regularity (or even any weaker form of robust regularity) in the behaviour of micro-systems would seem to make proper grasp of the properties possessed by the systems difficult or impossible.

Of course, there are good reasons to be chary of the notion of an unspeakable realm: we would be ill-advised to go so far as to suggest a quasi-Kantian realm of the noumenal. As Wittgenstein perspicaciously remarked, `A nothing would do as well as a something about which nothing could be said' \citep[\S 304]{wittgenstein:investigations}. So we should note that the unspeakability of \textit{our} micro-level is considerably mollified: we can certainly say \textit{some} things about its inhabitants, albeit not as much as we should like from the point of view of the traditional realist descriptive project (we perforce lack a complete and detailed dynamical theory, or theories, for instance). In particular, we can know that various micro-systems, of various types, exist: after all, these items can be isolated and experimented upon; we know that we are built from electrons and quarks, for example. We can know about these systems indirectly, by our experiment and causal interaction with them; and our fundamental micro-theory (which will, of course, always be apt for revision in its details) will provide the so-called sortal concepts under which these systems (or at least some of them) will fall, e.g., `electron', `neutrino', `quark', `mode of the electromagnetic field'. What we lack---what is unspeakable, for there is not much to say---is any \textit{detailed} story about these items and how they behave.

\subsubsection{Objectivity and the classical level}

Having presented an outline of, or framework for, an ontology we may at this stage remain agnostic on further details that might be pursued: it seems enough has been done to illustrate the kind of ontology that might best suit the quantum Bayesian's needs. But while remaining agnostic on some of the details of the proposed ontology, there are some pertinent questions one can pursue further, in particular, the question of what kind of relation there might be between the lawless micro-level and the relatively well-behaved macroscopic or classical level. The question of what might count, precisely, as the macroscopic or classical level is no doubt deserving of careful discussion, but we have sufficient pointers  already for a perfectly adequate rough-and-ready characterisation. The macroscopic or classical level will be a level of objects which \textit{do} have unproblematically stateable truths about them\footnote{Recall, we already employed this criterion when discussing above what gets to be an apparatus.}, a level of objects regarding which we \textit{can} make (more or less approximate) true generalisations; perhaps even roughly law-like ones in certain circumstances; perhaps generalisations which follow roughly the line of classical Newtonian physics. These are objects that we \textit{don't have to} invoke quantum mechanics to deal with, if we don't want to; if we don't delve too deeply. It should be no surprise or mystery that such a level of objects exists: they include the familiar physical objects which surround us and with which we are in constant interaction, day and night. We know that these objects are pretty well behaved and have a large numbers of stateable physical truths about them.

Now we can ask: how do the facts at the (roughly) classical macro-level relate to those at the micro-level? Although our model ontology is light on details, we can at least say this: the relation can be no stronger than supervenience of the macroscopic truths on the microscopic ones; and it might be weaker\footnote{If truths of domain $A$ are said to supervene on truths of domain $B$ then this means that there can be no change in the $A$-truths without a change in the $B$-truths, but not necessarily conversely. This basic definition can be elaborated in various ways to capture distinct notions: for example, one might have in mind change over time within a single model of the world (e.g., if all the $B$ facts in a given model remain the same over some interval of time, then there can be no change in the $A$ facts during that time) or one might have in mind changes between different models (if two models of the world---two different possible worlds---are the same with respect to all the $B$ facts, then they will also be the same with respect to the $A$ facts). Exactly which form of supervenience relation might be apt for our case is underpsecified by our model ontology, so we shan't pursue it further}. It can be no stronger because this would require some form of theory reduction between the two levels, minimally, general statements regarding micro-level facts which would imply when and where various macro-level truths and generalisations hold. But given that there can be no theory about the micro-level and no true generalisations about the behaviour of micro-systems, this is clearly impossible. The reason that the relation might be even weaker than supervenience is that it could be that when various micro-systems interact, they, taken as a whole composing some larger macro-object, can possess what are sometimes called \textit{metaphysically emergent} properties. These are properties that a composite can possess but which its components cannot; and which are not conferred on it by the properties possessed by its components and the laws (if any) which they obey. The emergent properties are added on top: one imagines all the micro-level properties of the world being specified and then one having to add these further truths to the mix: specifying all the micro properties isn't enough. While such emergent properties might in some instances be consistent with supervenience \citep{horgan:superduper}, they might also very well not be. So although we must insist that macro-objects are composed of the micro-systems and so are not ontologically autonomous objects (if one took the micro-systems away, then you'd perforce no longer have any macro-objects) it might be that there are some autonomous \textit{facts} about the macro objects; it might not be the case that the four dimensional pattern of truths, such as they are, regarding the micro-level is sufficient to fix the pattern of macroscopic truths.

An analogy that might prove helpful is with the doctrine of epiphenomenalism in the philosophy of mind. According to epiphenomenalism, mental goings-on both genuinely exist and are distinct from physical goings-on, yet they don't act on the physical goings-on; they just bob along on top. There is no claim of reduction of the mental to the physical, but neither of consequent elimination of the mental. Instead the idea is that mental happenings are a somewhat incidental (apart from to themselves!) by-product of the physical behaviour of the brain and central nervous system: something merely thrown up by the physical goings on; pleasant and interesting enough to enjoy, no doubt, but secondary to the level (the neurological) where the real action is. A froth thrown up on top of a deep sea rife with shifting currents. We can think of the relation between micro-level and macro-level in the quantum Bayesian picture as having something of this character: from the roiling, unspeakable mess of microscopic events is thrown up the level of relatively enduring, well-behaved and characterisable macroscopic objects. In contrast to the epiphenomenalist picture, though, which would locate causal laws and influences at the lower (neurological, physical) level, in our quantum Bayesian picture, the only place where anything vaguely law-like could be found is in relations between goings-on at the higher (that is the roughly classical, macroscopic) level.

This suggests the next question: how can we have complete lawlessness at the fundmamental physical level yet the possibility of lawfullness, or at least the possibility of pretty good true generalisations, at the macroscopic level? Does this possibility make sense? I think it does, on a number of counts. To begin with, it seems quite possible that unruliness at the fundamental level can simply wash out to allow useful (perhaps somewhat approximate) generalisations at a higher level: think of kinetic theory and thermodynamics, for example. It even seems quite intelligible that \textit{exact} laws could hold at the higher level on top of lawless underpinnings, irrespective of one's detailed view of laws. Thus if one were broadly Humean about laws, holding that all that is important for lawhood is the obtaining of particular kinds of regularity in the phenomena, then it could simply be that at the higher level such regularities \textit{do} obtain, albeit that there is wild unruliness beneath. Or perhaps one holds the view that laws are given by the obtaining of special higher-order `necessitation' relations between physical properties \citep{armstrong,dretske:laws}. Again, there seems to be nothing to rule out these relations obtaining between properties of the higher-level classical objects, but not between those of objects lower down; and the regularities at the higher level that these relations might be supposed to entail can indeed be in place, just as much as for the Humean. Finally we should note that this kind of challenge is one which already faced Cartwright's position: in her view, there is interaction between the causal powers of physical objects, giving rise to lawlikeness in some domains but not universally. Our consideration is just a special case of this in which all the domains in which laws hold are higher than the micro-level. Her response to the concern is simply to note that it could well be the case that the world \textit{just is} arranged in such a way that we do get consistency between the behaviour in these different domains, some governed by some laws, others by others, some not at all \citep[p.33]{cartwright:dappled}. Consistency is possible; and indeed if the world is composed of systems having causal powers which only sometimes give rise to lawlike behaviour in various restricted domains---a patchwork of laws and elsewhere unruliness---then, trivially, inconsistency is \textit{impossible}, logically so, so one doesn't need to worry about it.

\subsection{Troubles with explanation}\label{explanation}

If called upon, then, the quantum Bayesian seems able to present an intelligible ontology to underly their position. More worrying for the view is the question of how they fare with the possibility of explanation in the physical realm.

It seems incontrovertible that we require our physical theories to be able to provide us with explanations of various kinds: explanations of why things are as they are, or of how things work, or of how the processes going on at one level of organisation relate to those at another, for example. This, after all, is one of the main reasons why we go in for the business of science in the first place. The requirement that science must (by and large) involve the provision of explanations (amongst the many other things it might be called on to do) is one of the sticks with which one traditionally beats the instrumentalist. For there is a big difference between explaining something and predicting it (as we emphasise in traditional criticisms of the Deductive-Nomological approach to explanation, for example) yet the instrumentalist view of science would reduce theories to black boxes which merely spit out predictions and are incapable of furnishing explanations. We have already noted that the quantum Bayesian view is not an instrumentalist one: it does not maintain (indefensibly) that the business of science is merely prediction, not explanation; nor, more weakly, does it maintain this view restricted to the domain of microphysics alone. But for all that, it is hard to see how, if the quantum Bayesian approach to quantum mechanics were correct, we could have the kinds of explanation involving quantum mechanics that we certainly do seem to have.

The seeds for this worry were in fact sown earlier, in our discussion of the solipsism charge against quantum Bayesianism (Section~\ref{not solipsism}). There we noted that the quantum Bayesian position is not reductionist: given that quantum mechanical statements typically do not attribute properties to objects (certainly not those statements involving state assignments, in any case) they will not be apt to serve as a base to which other kinds of physical statements---about classical level objects, say---can be reduced. But one need not belong to the brigade of imperialists about physics---those who believe that \textit{all} scientific theories of whatever domain can, in a strong sense, ultimately be reduced to physics---to be concerned that we may have gone too far here. Blanket reductionist claims are implausible, but surely we can and do expect suitably modest flavours of reduction when we seek to explain why macroscopic or classical level objects have the kinds of physical properties that they do in virtue of the properties their constituents possess?

The examples need not be terribly \textit{recherch\'{e}}. Think of such bread and butter examples as explaining why matter is stable given that it is composed of charged particles, or of explaining the thermal and electrical conductivity properties of solid matter. These are amongst the most basic, but important, explanations that we think of quantum mechanics as providing: they count as great explanatory successes. But how, if quantum mechanics is not to be construed as a theory which involves ascribing properties to micro-objects along with laws describing how they behave, can we account for this explanatory strength?

Thus think of the question of why some solids conduct and some insulate; why others yet again are in between, while they all contain electrons, sometimes in quite similar densities. To answer this question, we need to talk about the behaviour of charge carriers in bulk solids and what influences it. It presumably will not do to be told how, for example, by conditioning on data, independent agents who have certain beliefs about the constituents of sodium  might come to have high degrees of belief that matter having that structure would conduct. This might be revealing if such reflections on beliefs could be read as a circumlocutory way of talking about facts about sodium in virtue of which one's beliefs would be justified; but in the quantum Bayesian picture, that is disallowed.

The central point is this: Ultimately we are just not interested in agents' expectation that matter structured like sodium would conduct; we are interested in \textit{why in fact it does so}. The normal quantum mechanical discussion proceeds by reflecting on how the periodic structure of ions in a solid influences the allowed electronic states, opening up gaps (\textit{band gaps}) in the energy spectrum of the allowed momentum states. Then one considers the location of the Fermi surface: the surface in (crystal) momentum space below which all the filled states lie. If this surface cuts a band of allowed states, the material will be a conductor as there are higher energy states for electrons to move into on application of a potential. If, however, the surface falls into a band gap between allowed states then the material will either be an insulator or a semi-conductor depending on the size of the gap to the next lot of allowed states. Now how much of this talk would remain as explanatory if we were to move to the quantum Bayesian position where there are \textit{no facts about what the states of electrons in a solid even are}?\footnote{Two points to note: the kinds of examples I am concerned with here are not, I take it, question begging against the quantum Bayesian, as, for example, a demand `How do you \textit{explain} the half-life of uranium if not by objective probabilities derived from objective states?' would be. The examples do not seem to turn centrally on probabilistic notions, but rather arise when having particular states for systems plays an explanatory role of some sort. Secondly, the instrumentalist is in bad shape with their programmatic silence about properties of systems at the microlevel and their insistence on prediction alone, but in some ways the quantum Bayesian is in \textit{worse} shape given that there aren't even any facts, for them, about what predictions from the theory would be right, given the facts about the world.}

A closely related concern is raised by Hoefer regarding Cartwright's `capactities first' based picture of science \citep{hoefer:fundamentalism}. Suppose we consider the simplest realistic quantum mechanical system of all, a hydrogen atom. We have two components, a positively charged proton and a negatively charged electron. We take these objects to be endowed with certain causal powers, whose joint interaction in certain circumstances will, in Cartwright's view, give rise to regular lawlike behaviour; behaviour that would be captured by the Schr\"{o}dinger equation for a single electron in a Coulomb potential. Her thought is that the capacities of the objects are basic and the Schr\"{o}dinger equation behaviour derivative from them. But basic in what sense? Could we just drop the Schr\"{o}dinger equation and work alone with the capacities borne by proton and electron? Showing how, in their interaction, these capacties in fact allow the formation of stable, non-radiating atoms? Certainly not. There is no hope of such an approach succeeding, for, as Hoefer points out, we would need to have independent grasp of what the capacities of the objects were and \textit{then} be able to show how, under the right circumstances, they gave rise to the possibility of stable atomic configurations. But of course we have no such independent access to the capacities. The only kind of detailed access to them that we have is \textit{via} the generalisations true of them: in this case, \textit{via} the Schr\"{o}dinger equation. The point can be put in terms of an explanatory defecit: if we insist on the capacities-first picture, then the only explanation we have of why stable hydrogen atoms are possible is this: `Protons and electrons are endowed with the capacity to form stable atoms under certain conditions'. This is laughably meagre explanatory fare (close to a `dormitive virtue'-style explanation) when compared to the richness that follows when one begins with the thought that these systems are governed by the Schr\"{o}dinger equation and proceeds to explore quantitatively as well as qualitatively exactly how and why stable conditions of various kinds arise.

The explanatory defecit problem seems to occur in the quantum Bayesian's picture too. Returning to our conductivity examples: what can we say about why metals conduct if we eschew quantum mechanics as being in some sense a descriptive theory? Well, we can at least say that a sample of metal will contain a large number of electrons: we know that these are charged and apt to be accelerated when a field is applied; and given that these electrons in the solid flow when a voltage is applied, they must be free to some degree to move around in the sample. This seems about the limit of the descriptive claims about the charge carriers and their behaviour that we can make if we remain within the quantum Bayesian's limits. When we are disallowing the possibility of any descriptive theory below this kind of level of generality it seems that there is little that we can say about why macroscopic objects have the properties they do, other than saying that they are composed of things which are disposed to give rise to this kind of behaviour under certain circumstances (the shadow of `dormitive virtue' explanation again). But then look how much more one gets when one turns to quantum mechanics more normally construed! Interestingly, it turns out that the model where one assumes that the electrons in a solid are completely free to move around (forming a kind of `gas' permeating the lattice of ions) actually fits the qualitative features of many metals pretty well, but in order to explain \textit{why} that assumption is a good one---why aren't the electrons scattered by the ions, for example?---and to explain other anomalies (e.g., specific heats, the Hall effect coefficients and magnetoresistance) \textit{and the very existence of insulators and semiconductors}, then one needs a detailed set of quantum mechanical claims about what is going on in the solid, as outlined above.

So the challenge is this: it looks like the quantum Bayesian faces an explanatory defecit. It seems that we do have very many extensive and detailed explanations deriving from quantum mechanics, yet if the quantum Bayesian view were correct, it is unclear how we would do so. For it would deny what seems to be a crucial part of the functioning of these kinds of explanations, namely, that quantum mechanics makes claims about the properties of micro-systems and describes how they behave.      

Now it should be conceded that this perhaps amounts only to a \textit{prima facie} challenge, for the quantum Bayesian could always explore the possibility of endorsing or developing some heterodox account of explanation which could show why quantum mechanics would still count as providing us with explanations notwithstanding its non-descriptive character. Perhaps something along the lines of Cartwright's so-called \textit{simulacrum account} of explanation \citep[Chpt. 6]{cartwright:hlpl}, whereby to explain some phenomenon is to provide a model which fits it into the theory, might be a suitable starting point. But until some such approach is developed, the question of explanation does seem to raise considerable difficulties for the quantum Bayesian's conception.

\subsection{Subjective probabilities}\label{subprobs}

Our final locus of concern is whether subjective probabilities can really do the job in quantum mechanics. It is sometimes baldly asserted that they cannot: that probabilities in quantum mechanics \textit{must} be objective: what could properties such as half-lives for radioactive decay (for example) \textit{be} but objective physical properties? These aren't a matter of what one thinks, but of how things are! Or so one might be inclined to insist. However such assertions clearly carry no weight against the quantum Bayesian, as they are question begging. For the quantum Bayesian takes themselves to be in possession of a perfectly adequate account of probabilistic reasoning in quantum mechanics. So our route will be to begin by highlighting a certain uncomfortable oddity that the quantum Bayesian's position commits them to, before raising a more systematic doubt for the conception.  

\subsubsection{A Quantum Bayesian Moore's Paradox}

G.E. Moore highlighted an interesting oddity in the logic of belief assertions. It seems plainly wrong, even paradoxical, to assert a sentence like 
\begin{itemize}
\item `It is raining, but I don't believe it', or 
\item`It is raining, but I believe that it is not raining'. 
\end{itemize}
Such utterances seem to hang uncomfortably between being straight contradictions (`It is raining and it is not raining') and being empty strings of words which fail to add up to saying anything. But following Moore (cf. \citet{austin:words}) the orthodox view came to be that one can see these sentences as making sense; and can even imagine circumstances in which their utterance might express a truth: circumstances, that is, in which I am right in what I say about the weather (`It is raining') but where I happen to have the wrong belief about it. Such circumstances might be highly atypical---if sincere, my statements about the weather and my beliefs about it will usually line up---but they do seem possible. Straight contradiction is avoided, runs the thought, because the first half of the sentence is used to state something about the world external to myself, while the second half is used to state something about me: the two halves aren't quite talking about the same thing. Then what is wrong about these \textit{Moore's paradox} sentences, what makes them uncomfortable and paradoxical, is not that they are nonsense or contradiction, but that they force one to \textit{violate the rules for the speech act of sincere assertion}. To a first approximation (\citet{austin:words,searle:acts}), in order to make a sincere assertion, one needs to believe what one is going to assert; and that one believes it will be \textit{expressed}---although not stated---by one's assertion. The trouble with the Moore's paradox sentences is that they subvert these rules: in uttering such a sentence, one would be denying that one had the belief that is a pre-requisite for validly making the initial assertion. Thus they can never be used to make \textit{licit} assertions---herein lies their \textit{pragmatic paradox}---even if they are meaningful and there are circumstances in which they could express truths. (While this is the orthodox view of these sentences, it is worth noting that there remains a minority view stemming from Wittgenstein's discussion \citep[{II}.x]{wittgenstein:investigations} (cf. \citet{heal:moores}) which maintains that the sentences involve a semantic, not just a pragmatic, contradiction.)   

Now oddly enough, the quantum Bayesian seems to be committed to the \textit{systematic} endorsement of utterances which are very closely analogous---and similarly uncomfortable---to Moore's paradoxes. These arise when one considers pure state assignments.\footnote{That there might be some oddity for the quantum Bayesian's position around this area was first suggested to me by Howard Wiseman at a conference in Konstanz in 2005. The following argument and the connection to Moore's paradox can be seen as an attempt to develop the concern and make it more precise.}

Pure states are absolutely central to the theoretical structure and standard presentations of quantum mechanics, although perhaps somewhat less so to the practice of experimentalists. For the subjective Bayesian, they represent a special case; for not only will their assignment involve making a number of probabilistic claims, but will also involve a claim of \textit{certainty}. If I assign a pure state to some system---an eigenstate of some observable---then I am \textit{certain} that if that observable were to be measured (using a good measuring device), the result would be thus-and-so. Pure state assignments are black and white in a way that mixed state assignments are not. Because of this, one might be inclined to think that pure states are less subjective than mixed states---after all, it is perhaps easier to imagine how different agents, knowing different things, could assign different mixed states to one and the same system; and one might imagine that as these agents gather more data they will refine their belief states to narrow down on the actual pure state the system possesses. For non-extremal states, different state assignments can be explained away as arising from access to different data; not so any differences in pure---extremal---states. However Caves, Fuchs and Schack maintain, quite correctly, that for the quantum Bayesian, pure states can and ought to be understood in just the same---fully subjective---way as mixed states \citep{fuchs:what,cfs:certainty}. Any state assignment, pure or mixed, is a purely subjective matter, not one determined by the facts; otherwise, their central claims about the nature of measurement and non-locality do not go through. 

Nonetheless, one might still be puzzled by the subjectivity of a pure state assignment. In particular, one might be puzzled about how one could be certain that some particular outcome will obtain, when---by the quantum Bayesian's lights---there is no fact about what that outcome will be (no fact that makes the state corresponding to that outcome the right one). Don't I need something to explain why that outcome is going to be the way it is, something which would justify my certainty about it? Caves, Fuchs and Schack argue not \citep{cfs:certainty}; maintaining that the subjective Bayesian picture has the resources to answer this kind of concern; and on the face of it they seem right.     

They begin by clarifying what might be supposed to be in need explanation. Imagine that an agent assigns the pure state `spin-up in the $z$-direction' to a spin-half system and proceeds to perform a long sequence of non-disturbing $z$-spin measurements on the system, getting outcome `up' each time. One might think that in the absence of an actual fact that the system was in the state `spin-up', it would be both surprising and in need of explanation that this particular sequence should be observed. But surprising for whom, they enquire? \textit{Not} for the agent in question: on the contrary, given that he had assigned a pure state and hence was certain, he would have been surprised (to say the least!) had that sequence \textit{not} been observed. And what would be required to explain the sequence, from his point of view? Nothing, given that it would be of no surprise to him that it occurred---in reaching the pure state assignment in the first place, all the factors that he deemed relevant to what was going to happen would have been considered; if there were any pertinent gaps in the story about what was going to happen then a pure state would not have been assigned. Finally \textit{the agent's certainty} about the outcome is to be explained by citing those factors which he considered in reaching the original assignment. It is not obvious that more is required.

\sloppy This defence of the subjectivity of pure state assignments might be aided further by appeal to a distinction drawn by A. R. White between the certainty of people and the certainty of things \citep{white:certainty,white:modalthinking}. This follows a grammatical distinction between personal and impersonal forms of certainty statements: `I/You/Freddy...is/are certain' versus `It is certain'. White proposes that some\textit{thing} is certain if circumstances (facts about the world) exclude the possibility of its being othwerwise, whereas some\textit{one} is certain (that $p$) if they are settled that $p$ because they do not entertain the possibility of its being otherwise---perhaps because, but not necessarily because, they think that \textit{it} is certain that $p$.\footnote{White has in mind a non-epistemic sense of \textit{possible that} and hence a non-epistemic sense of certainty of things. That is, he does not seek to analyse the certainty of things in terms of the certainty of people, or of suitable populations of people. This might be a matter for debate, but does not seem crucial for current purposes.}

White goes on to argue that the certainty of people and the certainty of things are logically independent, thus: I can be certain that $p$ without \textit{it} being certain that $p$ (I can be certain that $p$ when it is just false that $p$---maybe I have the wrong expectations---\textit{a fortiori} I can be certain that $p$ without \textit{it} being certain that $p$) and \textit{vice versa}, \textit{it} can be certain that $p$, yet I'm not certain that $p$ (perhaps because I have various false beliefs, or irrelevant worries, or just don't know all the facts). Moreover, White argues that an utterance `It is certain' does not state, but only \textit{expresses} my certainty (compare above: `It is raining' does not state, but expresses my belief, hence the tension---but non-contradiction---in the Moore's paradox sentences). Finally, somebody who is certain that $p$ \textit{need not} logically (even if they typically might) think that \textit{it} is certain that $p$; and again \textit{vice versa}. For example, consider a frenzied gambler who is \textit{certain} that it will be red next time. He may have this belief without having to think that there is any mechanism in the roulette table or elsewhere which is going to entail that the ball will land red, that makes \textit{it} certain: he just has this conviction about the outcome: \textit{he} is certain, \textit{toute courte}. Or in the other direction, consider an obsessive worrier who definitely thinks that it is certain that $p$, yet whose timorous nature nonetheless precludes that producing in him certainty that $p$.

All this is grist to the quantum Bayesian's mill. If certainty of people and certainty of things are logically independent then one can be certain of something without anything guaranteeing that things will be so. Specifically, \textit{one} can be certain that the outcome of measurement will be spin-up in the $z$-direction (for example) while \textit{it} is not certain that the outcome will be spin-up: no fact entails it. So far so good.

What now if we ask: can one be certain that $p$ when one doesn't know that \textit{it} is certain that $p$? Again, this will be allowed; in fact it is entailed by the previous case. But now consider an apparently closely related statement: 
\begin{itemize}
\item Can one be certain that $p$ when one knows (or believes) that \textit{it} is not certain that $p$?
\end{itemize}
Now we seem to have trouble.

For notice that a proponent of quantum Bayesianism, who is self-conscious about their position and their understanding of quantum states, must be happy to assert sentences like:
\begin{itemize}          
\item `I assign a pure state (e.g. $\ket{\up_{z}}{}$) to this system, but there is no fact about what the state of this system is.'
\end{itemize}
That is, they will be committed to asserting sentences like:
\begin{itemize}
\item \textbf{QBMP:} `I am certain that $p$ (that the outcome will be spin-up in the $z$-direction) but \textit{it} is not certain that $p$.'
\end{itemize}
And isn't such a sentence paradoxical? In much the same way that Moore's sentences are? The two halves of the assertion are in some kind of conflict: this is what we may call a quantum Bayesian Moore's paradox. It doesn't suffice that it is quite in order to assert that one is certain that $p$, when \textit{it} may not be certain that $p$; for given the nature of their position, the quantum Bayesian must in addition be happy to assert at the same time \textit{both} that they are certain that $p$ \textit{and} that \textit{it} is not certain that $p$; and this is quite another thing.

Caves, Fuchs and Schack maintain:
\begin{quote}
It might...be argued that an agent could not be certain about the outcome ``Yes" without an objectively real state of affairs guaranteeing the outcome, i.e., without the existence of an underlying instruction set. This argument, it seems to us, is based on a prejudice. What would the existence of the instruction set add to the agent's beliefs about the outcome? \citep[p.271]{cfs:certainty}
\end{quote}
This seems right: the prejudice in question is based on a failure to appreciate the distinction between certainty of people and certainty of things; as Caves, Fuchs and Schack remark, when associated with the assignment of a pure state by an agent, `Certainty is a function of the agent, not of the system' \citep[p.258]{cfs:certainty}. Yet identification of that prejudice is not the full story. One might grant that addition of the instruction set---the fact determining what the outcome will be, or the fact determining what the real state is---would not add anything to an agent who already assigned a pure state; but conjoining the explicit \textit{denial} that there is any such instruction set or relevant facts---something the quantum Bayesian must be prepared to do as a matter of course---cannot be met with equanimity. It sets up an unbearable tension. Why isn't the first half of the assertion undermined by the second half? How can the agent's certainty about an outcome reasonably be maintained when it runs concurrently with an explicit denial that the outcome is certain? Isn't the agent simply convicting themselves as irrational?

Three points to note. First, the difficulty we have broached here is one which would seem to affect subjective Bayesianism in general, not just as it finds application to quantum mechanics; at least whenever extremal probability assigments feature. The point about the quantum example is that the pure state assignments make up a very important class of cases which can't just be ignored as oddities; at least not without further argument. Secondly, we can imagine other kinds of cases---not just subjective Bayesian ones---where sentences like \textbf{QBMP} might come to the fore: Imagine that our frenzied gambler comes to a moment of lucidity, realising his thoughts: `I am certain that it'll be red; but it's not certain that the outcome will be red'. We can imagine him nonetheless sticking with his conviction; and would be happy thereby simply to grant him as irrational in this instance. That such thoughts are intelligible and consistent with the concept of certainty follows once we allow the logical independence of certainty of people and things. What is special about the quantum Bayesian case is that, again, the occurence of these paradoxical sentences isn't just an occasional oddity which can be ignored---as the gambler case might be---rather, the phenomenon is absolutely generic. It arises whenever one finds a quantum Bayesian who is happy to assign pure states and is also explicit about what their understanding of the quantum state is. Finally, while noting their similarity, I have not said exactly what the relation between the standard Moore's paradox sentences and the quantum Bayesian versions involving certainty are. Are they perhaps at root identical? If the equation `` `I am certain that $p$' = `I believe that: it is certain that $p$'" were true then they would be; but the equation seems false (cf. \citet[p.76]{white:modalthinking}). This is a matter for further investigation; as is the question of how the minority view that Moore's paradoxes are genuinely, not just pragmatically, inconsistent might transform the quantum case.

What are we to make of all this? The quantum Bayesian is unlikely to concede that this \textit{nouveau} Moore's paradox is an insuperable objection to their position: there seems to be a fair amount of wriggle room left to them. For example, one might suggest that the two halves of the pertinent sentences concern different levels of discussion; that the second claim---`it is not certain'---is made at a meta-level of philosophical analysis; is said, perhaps, in a different \textit{tone of voice}, from the ground level `I am certain'; and thus the tension is defused. Or one might point to apparently similar phenomena in other areas of philosophy. Consider for example error theorists or non-cognitivists about ethics---those who hold that there are no moral facts. Such theorists would nonetheless be happy to utter such apparently paradoxical sentences as `Stealing is wrong, but it's not true that stealing is wrong'; and this isn't generally taken to be an objection to their position. The reason for this, presumably, is that these theorists have (or purport to have) a detailed story to tell about the \textit{purpose} of moral discourse; and this is a story in which that purpose may be served without there being any moral truths. Perhaps the quantum Bayesian could similarly elaborate on how there can be a role for personal certainty within our intellectual economy which is insulated against the absence of any impersonal certainty.

Such moves might be made; but I am nonetheless inclined to see the quantum Bayesian Moore's paradox straightforwardly, as a significant difficulty for the quantum Bayesian. This is because the trouble it raises seems of a piece with---is perhaps a direct symptom of---a more general worry to which we now turn; a worry that in the quantum Bayesian setting, something has gone wrong in the relation between the reasons one can have and one's beliefs; in how one's reasons could be good bases for action.              

\subsubsection{The means/ends objection}

At a great level of generality, it seems reasonable to insist that when considering some domain of enquiry, there should be an appropriate match between the \textit{means} of the enquiry and its \textit{end}: what it seeks to achieve. What is puzzling about the quantum Bayesian position is how, if its premises are granted, its means could be expected to achieve its ends. What are these ends? 

Well, the quantum Bayesian, recall, is at heart a realist about physics; is one who believes in the existence of a mind-independent world of microscopic---and macroscopic---goings on; but is also one who concedes that there is a limit to what we can describe of this world. In their view, our best theory of the very small (quantum mechanics) is a theory which has to be understood pragmatically, as a way of \textit{dealing} with a world which refuses direct description at the fundamental level. We might, therefore perceive two distinct ends: one of finding out how the world is; the other the pragmatic business of coping with the world. Let us focus on the latter of these, as the logically weaker of the two.

Now what are the means? The familiar business of setting up experiments, collecting data and drawing inferences on the basis of this data. When we get down to our best physical theory, where probabilities rule, this will be the business of updating our subjective probabilities on the basis of the data we've generated. The puzzle is this: if there are only subjective probabilities, if gathering data does not help us track the extent to which circumstances favour some event over another one (this is the denial of objective single case probability), then why does gathering data and updating our subjective probabilities help us do better in coping with the world (if, that is, it does so)? Moreover, why should it be expected to? Why, that is, should one even bother to look at data at all? It's not as if it's going to guide us in what the world will throw at us; it just leads us to a different subjective probability distribution. An unexplained gap opens up between the means of the enquiry and its ends. Put in terms of reasons and beliefs: if one's reasons derive from the experiments one has performed then it is unclear how these could provide \textit{good} reasons for belief that such-and-such is to be expected, or good reasons to act in such-and-such a way. For example, good reasons to be certain that $x$ will happen are typically reasons for thinking that \textit{it} is certain that $x$ will happen; but the latter are never available by the quantum Bayesian's lights.

There is an immediate reply to this, of broadly Darwinian stripe. That is: We just \textit{do} look at data and we just \textit{do} update our probabilities in light of it; and it's just a brute fact that those who do so do better in the world; and those who don't, don't. Those poor souls die out. But this move only invites restatement of the challenge: \textit{why} do those who observe and update do better? To maintain that there is no answer to this question, that it is just a brute fact, is to concede the point. There is an explanatory gap.

By contrast, if one maintains that the point of gathering data and updating is to track objective features of the world, to bring one's judgements about what might be expected to happen into alignment with the extent to which facts actually do favour the outcomes in question, then the gap is closed. We can see in this case how someone who deploys the means will do better in achieving the ends: in coping with the world. This seems strong evidence in favour of some sort of objective view of probabilities and against a purely subjective view, hence against the quantum Bayesian.

Throughout the course of our discussion, I have been careful to highlight and discount those objections, or putative objections, to the quantum Bayesian position which simply beg the question against subjective probabilities. Isn't the means/ends objection just sketched merely another one of these? I think not. Perhaps something like these means/ends concerns does form part of the background to many a dogmatic rejection of subjective probabilities, but the reflections themselves are not dogmatic. The form of the argument, rather, is that there exists a deep puzzle if the quantum Bayesian is right: it will forever remain mysterious why gathering data and updating according to the rules should help us get on in life. This mystery is dispelled if one allows that subjective probabilities should track objective features of the world. The existence of the means/ends explanatory gap is a significant theoretical cost to bear if one is to stick with purely subjective probabilities. This cost is one which many may not be willing to bear; and reasonably so, it seems.

\section{Conclusions}

We began by exploring the quantum Bayesian position in some detail and seeing that it could be defended from certain objections that had been levelled against it. It is clear that once quantum Bayesianism is properly understood, it does not admit of the charge of solipsism nor yield to the challenge that it is straightforwadly instrumentalist. Perhaps the two most important features developed in the presentation were a) the elaboration of the philosophical setting of the quantum Bayesian position: how it seeks to retain a realist view of physics whilst admitting limits to what can be described; and b)the elucidation of how the quantum/classical divide functions in this setting: how a shift in what gets treated as quantum or as classical is natural and unproblematic in this approach; most of all, how a shift from treating an item as classical to treating it as quantum is not associated with a shift in the ontological status of the object. We saw that it is perfectly permissible to assign a non-classical quantum state to macroscopic objects about which there are nonetheless determinate classical level facts.

These are positive features, but we may not conclude that quantum Bayesianism is unproblematic. In our consideration of three substantive questions that the position faces, we saw that the first---the question of ontology---could be managed. While the ontological picture that seems most naturally to go along with the quantum Bayesian proposal of a world unspeakable at the fundamental level is not a standard one, it does not seem unintelligible, nor intrinsically objectionable. In fact it drew on what are familiar elements in current philosophy of science: a Cartwrightian picture of the causal powers or capacities of objects as basic and universal fundamental laws as absent. However, we saw that on the issue of explanation and the issue of subjective probability significant problems present themselves. Quantum Bayesianism, as it stands, faces the explanatory defecit problem: it is unclear how what \textit{is} explanatory could be so. Regarding probabilities, we found that the quantum Bayesian is committed to endorsing objectionable Moore's paradox-type assertions and saw that their position has the unfortunate weight to bear of a worrying gap between the means and ends of the enquiry. It may be possible to resolve these problems; the challenge for the quantum Bayesian is to do so.

\section*{Acknowledgements}

This work was supported by a research leave award from the UK Arts and Humanities Research Council. I would also like to thank the Department of Philosophy at the University of Leeds for further research leave which allowed this work to be pursued. For most useful discussion I would like to thank Carl Caves, R\"{u}diger Schack and, especially, Chris Fuchs. Also: Jon Barrett, Harvey Brown, Ari Duwell, Joel Katzav, Joseph Melia (in particular), Veiko Palge, Rob Spekkens, David Wallace, Robbie Williams and Howard Wiseman. 

\newpage

\end{document}